\documentclass[%
amsmath,amssymb,
aps,
prfluids,
groupedaddress,
]{revtex4-2}

\usepackage{graphicx}% Include figure files
\usepackage{dcolumn}% Align table columns on decimal point
\usepackage{bm}% bold math
\usepackage{lipsum} % µ÷ÓÃ¿çÀ¸³¤¹«Ê½ºê°ü
\usepackage{float}
\usepackage{mathrsfs,amsthm,amsmath,amssymb}
\usepackage[colorlinks=true,linkcolor=blue]{hyperref}%
\usepackage{color}
\begin{document}

%\preprint{APS/123-QED}

\title{Machine-learning-based simulation of turbulent flows over periodic hills using a hybrid U-Net and Fourier neural operator framework}% Force line breaks with \\
\author{Yunpeng Wang$^{1,2}$}
\author{Huiyu Yang$^{1,2}$}
\author{Zelong Yuan$^{3}$}
\author{Zhijie Li$^{4}$}
\author{Wenhui Peng$^{5}$}
\author{Jianchun Wang$^{1,2}$}
\email{wangjc@sustech.edu.cn}
\affiliation{\small 1. Department of Mechanics and Aerospace Engineering, Southern University of Science and Technology, Shenzhen, 518055 China.\\
2. Shenzhen Key Laboratory of Complex Aerospace Flows, Department of Mechanics and Aerospace Engineering, Southern University of Science and Technology,Shenzhen 518055, China.\\
3. Harbin Engineering University Qingdao Innovation and Development Base, Qingdao 266000, China.\\
4. Department of Biomedical Engineering, National University of Singapore, Singapore, 117583, Singapore.\\
5. Department of Energy and Power Engineering, Nanchang Hangkong University, Nanchang, 330063, China.}
\date{\today}% It is always \today, today,

\begin{abstract}
Simulating massively separated turbulent flows over bodies is one of the major applications for large-eddy simulation (LES). In the current work, we propose a machine-learning-based LES framework for the rapid simulation of turbulent flows over periodic hills using a hybrid U-Net and Fourier neural operator (HUFNO) framework. The newly proposed HUFNO model integrates the strengths of both the convolutional neural network (CNN) and Fourier neural operator (FNO) in a novel way that the FNO is applied in the periodic directions of the flow field while the non-periodicity is handled by the CNN-based U-Net framework. In the numerical tests, compared to the original FNO and the U-Net framework, the HUFNO model shows a higher accuracy in the predictions of the velocity field and Reynolds stresses. Further numerical experiments in the LES show that the HUFNO framework outperforms the traditional Smagorinsky (SMAG) model and the wall-adapted local eddy-viscosity (WALE) model in the predictions of the turbulence statistics, the energy spectrum, the invariant characteristics of velocity gradients, the wall stresses and the flow separation structures, with much lower computational cost. Importantly, the accuracy and efficiency are transferable to unseen initial conditions, Reynolds number and hill shapes, underscoring its great potentials for the fast prediction of strongly separated turbulent flows over curved boundaries.

%\begin{description}
%\item[PACS number(s)]
%47.20.Ky, 47.20.Lz, 47.20.Ma
%\end{description}
\end{abstract}

\pacs{Valid PACS appear here}% PACS, the Physics and Astronomy
\maketitle
%\tableofcontents


\section{\label{sec:intro}Introduction}

Accurate and efficient predictions of three-dimensional (3D) turbulent flows are of significant importance in the fields of science and engineering \cite{Pope2000}. As a critical flow prediction technique, computational fluid dynamics (CFD) offers valuable insights into flow field information, particularly when experimental data are challenging to obtain. While traditional flow simulation methods continue to develop \cite{Yang2021,Duraisamy2019,Bin2024,Raje2025,Sagaut2006,Smagorinsky1963,Lilly1967,Deardorff1970,Germano1992,Mons2021,Rozema2022}, machine learning (ML)-based techniques for flow prediction have been widely proposed over the past decade, thanks to the rapid advancement of modern computing power and the accumulation of high-fidelity data \cite{Vinuesa2023,Li2024}. These efforts encompass a broad range of applications, including learning aerodynamic forces in the flow fields \cite{Wang2022,Mohamed2023,Tran2024}, modeling unclosed terms in the governing equations \cite{Wang2017,Yang2019,Xie2020,Duraisamy2021,Kurz2023,Xu2023}, wall modeling \cite{Vadrot2023,Zhou2024,Zhou2024a}, inferring missing information due to measurement constraints \cite{Buzzicotti2021,Li2023a,Zhuang2025}, flow field super-resolution \cite{Deng2019,Kim2021,Fukami2024}, and directly learning the temporal evolution of flow dynamics \cite{Li2020,Peng2022,Han2022,Peng2023,Mohan2019,Mohan2020,Nakamura2021,Bukka2021,Srinivasan2019,Yang2024,Yang2026,Wang2024,Raissi2019,Wang2020,Jin2021,Zhao2025}, etc. In these applications, directly simulating the flow field evolution has received increasing attention recently, considering that trained models can provide rapid evaluations of detailed flow information without solving the governing equations.

In the field of ML-based flow predictions, commonly adopted methods include the physics-informed neural networks (PINN) \cite{Raissi2019,Wang2020,Jin2021,Zhao2025}, the recurrent neural networks (RNN) and long short-term memory (LSTM)-based frameworks \cite{Mohan2019,Mohan2020,Han2022,Nakamura2021,Bukka2021}, and neural operator-based techniques \cite{Li2020,Tran2021,Li2022,Peng2022,Peng2023,Wang2024,Zou2025}. For instance, Raissi et al. proposed a PINN-based framework to solve general nonlinear partial differential equations \cite{Raissi2019}. Wang et al. proposed a turbulent flow network (TF-Net) based on a specialized U-Net architecture incorporating physical constraints in the context of two-dimensional (2D) Rayleigh-Bénard (RB) convection \cite{Wang2020}. Bukka et al. integrated proper orthogonal decomposition (POD) with deep learning (DL) to simulate the flow past a cylinder \cite{Bukka2021}. Han et al. adopted the CNN and LSTM-based frameworks to predict flow field evolution and fluid-solid interaction \cite{Han2022}. 

Although many neural networks (NNs) can learn the mappings between finite-dimensional Euclidean spaces for specific datasets, they often struggle to generalize across different flow conditions or boundary configurations \cite{Raissi2019,Li2020}. To address this limitation, Li et al. introduced an innovative Fourier neural operator (FNO) framework, which efficiently learns mappings between high-dimensional features in Fourier space, enabling the reconstruction of information in infinite-dimensional spaces \cite{Li2020}. Their proposed FNO model demonstrated remarkable accuracy in predicting two-dimensional (2D) turbulent flows. Following Li et al.'s work, numerous extensions and applications of FNO have been developed \cite{Tran2021,Li2022,You2022,Wen2022,Peng2022,Li2023d,Li2023,Peng2023,Li2023c,Peng2024,Wang2024,Zou2025}. Tran et al. proposed a factorized version of the FNO framework to enhance the efficiency of the FNO framework \cite{Tran2021}. Peng et al. incorporated an attention mechanism into the FNO framework, leading to improved statistical predictions and instantaneous flow structures \cite{Peng2022}. You et al. developed an implicit Fourier neural operator (IFNO) \cite{You2022}, to reduce the number of trainable parameters and memory requirements. Wen et al. proposed a U-Net enhanced FNO (UFNO), which outperformed both the original FNO and CNN frameworks in solving complex gas-liquid multi-phase problems \cite{Wen2022}. Recently, Peng et al. further applied the FNO in real-time simulation of 3D dynamic urban microclimate \cite{Peng2024}.

It should be noted that many of the existing NN methods are developed in the scenario of laminar flows or 2D turbulence problems. For 3D turbulence, the nonlinear interactions are fundamentally more complex than those in 2D turbulence. The increased model complexity also leads to higher memory demands and a greater number of NN parameters \cite{Li2022}. Mohan et al. introduced two reduced-order models based on a convolutional generative adversarial network (C-GAN) and a compressed convolutional LSTM (CC-LSTM) network in the context of 3D homogeneous isotropic turbulence \cite{Mohan2019,Mohan2020}. Peng et al. advanced the FNO model by incorporating a linear attention mechanism, leading to improved efficacy in the 3D homogeneous isotropic turbulence and turbulent mixing layers \cite{Peng2023}. More recently, Li et al. trained FNO and implicit U-Net enhanced FNO (IUFNO) models using coarsened DNS data of 3D isotropic turbulence and turbulent mixing layers \cite{Li2022,Li2023}. Their models can be viewed as surrogate models for the large-eddy simulation (LES) of turbulent flows. For wall-bounded flows, Nakamura et al. integrated a 3D CNN autoencoder (CNN-AE) with an LSTM network to predict 3D turbulent channel flow \cite{Nakamura2021}. Jin et al. developed Navier-Stokes flow nets (NSFnets) using the physics-informed neural network (PINN) framework to predict the solutions in small subdomains of turbulent channel flow \cite{Jin2021}. Recently, by enforcing the solid-wall boundary conditions, Wang et al. extended the application of the IUFNO framework to the LES of 3D turbulent channel flows at various Reynolds numbers, leading to better predictions compared to the traditional LES simulations \cite{Wang2024}. It is important to note that these studies on wall-bounded turbulence have only considered simple boundaries. However, the solid boundaries in real-world scenarios are more complex and often curved, potentially giving rise to strong flow separations, which increase the prediction difficulties.

In the present work, we focus on the strongly separated turbulent flows bounded by curved solid walls. Meanwhile, to account for the non-periodic boundary conditions often encountered in problems with complex geometries, we combine the convolutional neural network (CNN)-based U-Net framework and Fourier neural operator (FNO) in a hybrid approach, namely the hybrid U-Net and FNO (HUFNO) framework. It is worth pointing out that the HUFNO framework is fundamentally different from our previously proposed IUFNO model, since the FNO module in IUFNO is adopted in all three coordinate directions while the embedded U-Net IUFNO is only responsible for learning the residual small-scale dynamics \cite{Li2023,Wang2024}. In contrast, the FNO and U-Net are respectively responsible for the periodic and non-periodic flow dynamics in the current HUFNO framework. In the current investigation, we test the performance of HUFNO in the context of surrogate modeling for LES, considering that simulating all scales of turbulence for wall-bounded 3D turbulence at moderately high Reynolds numbers remains challenging for NN-based approaches. To incorporate the effect of curved boundaries, we adopted the turbulent flows over periodic hills as the benchmark problem. To the best of our knowledge, this represents the first attempt to develop surrogate LES models for strongly separated 3D wall-bounded turbulence in the presence of curved boundaries, with transferable accuracy and efficiency to unseen initial conditions, Reynolds number and hill shapes.

The rest of the paper is organized as follows. The governing equations of incompressible turbulence and traditional LES are introduced in Section \ref{sec:equ_les}, followed by a brief discussion on the solution strategies for the LES equations. The newly proposed HUFNO framework is introduced in Section \ref{sec:HUFNO}. In Section \ref{sec:post}, the performances of the proposed HUFNO frameworks are evaluated in the simulations of periodic hill turbulence, and compared against the DNS benchmark and the traditional LES results. Finally, a brief conclusion of the paper and some future perspectives are given in Section \ref{sec:conclu}.

\section{\label{sec:equ_les}Governing equations of turbulent flows and the large-eddy simulation}

In this section, the governing equations of incompressible turbulence are introduced, followed by a brief description of the traditional large-eddy simulation (LES) and the corresponding subgrid-scale (SGS) models.

The governing equations for the evolution of incompressible turbulence are given by \cite{Pope2000,Ishihara2009}
 \begin{equation}
  \frac{\partial u_{i}}{\partial x_{i}}=0,~~~ \frac{\partial u_{i}}{\partial t}+\frac{\partial u_{i} u_{j}}{\partial x_{j}}=-\frac{\partial p}{\partial x_{i}}+\nu\frac{\partial^{2} u_{i}}{\partial x_{j}\partial x_{j}}+\mathcal{F}_{i},
  \label{N_S}
\end{equation}
where $u_{i}$ is the $i$th velocity component, $\nu$ is the kinematic viscosity, $p$ is the pressure divided by the constant density $\rho$, and $\mathcal{F}_i$ accounts for any external forcing. Throughout this paper, the summation convention is used unless otherwise specified. Based on the velocity field, the strain and rotation rate tensors can be respectively defined by ${S}_{ij}=(\partial {u}_{i}/\partial x_{j}+\partial {u}_{j}/\partial x_{i})/2$ and ${\Omega}_{ij}=(\partial {u}_{i}/\partial x_{j}-\partial {u}_{j}/\partial x_{i})/2$.

To date, solving the NS equations using fine-grid DNS is yet impractical at high Reynolds numbers due to large range of scales involved \cite{Pope2000,Yang2021}. Compared to DNS, LES only simulates the dominant large-scale energy-containing motions using a coarse grid, and the subgrid-scale effects are modeled as functions of the resolved scales. \cite{Sagaut2006,Germano1992,Smagorinsky1963,Lilly1967,Deardorff1970}. By filtering the NS equations, the governing equations for LES can be obtained as \cite{Pope2000,Sagaut2006}
\begin{equation}
  \frac{\partial \overline{u}_{i}}{\partial x_{i}}=0,~~~ \frac{\partial \overline{u}_{i}}{\partial t}+\frac{\partial \overline{u}_{i} \overline{u}_{j}}{\partial x_{j}}=-\frac{\partial \overline{p}}{\partial x_{i}}-\frac{\partial\tau_{ij}}{\partial x_{j}}+\nu\frac{\partial^{2} \overline{u}_{i}}{\partial x_{j}\partial x_{j}}+\overline{\mathcal{F}}_{i},
  \label{fitered_N_S}
\end{equation}
with the overbar denoting a filtered variable calculated as $\overline{f}(\mathbf{x})=\int_{D}f(\mathbf{x}-\mathbf{r})G(\mathbf{r},\mathbf{x};\overline{\Delta})d\mathbf{r}$, where $f$ represents the physical quantity of flow field, here being velocity and pressure. G denotes the filter kernel, $\overline{\Delta}$ is the filter width and D represents the physical domain. The unclosed SGS stress $\tau_{ij}=\overline{u_{i}u_{j}}-\overline{u}_{i}\overline{u}_{j}$ represents the nonlinear interactions between the resolved and SGS motions. If the SGS stress can be modeled in terms of the resolved variables, the LES equations are closed.

A well-known model for SGS stresses is the Smagorinsky (SMAG) model \cite{Smagorinsky1963}, namely
\begin{equation}
 \tau^{A}_{ij}=\tau_{ij}-\frac{\delta_{ij}}{3}\tau_{kk}=-2C^{2}_{SMAG}\overline{\Delta}^{2}|\overline{S}|\overline{S}_{ij},
  \label{tauDSM}
\end{equation}
where $\overline{\Delta}$ is the filter width and $\overline{S}_{ij}$ is the filtered strain rate. The characteristic filtered strain rate is defined as $|\overline{S}|=\sqrt{2\overline{S}_{ij} \overline{S}_{ij}}$. Through some theoretical arguments for isotropic turbulence, one can deduce that $C_{SMAG}=0.16$ \cite{Smagorinsky1963,Lilly1967,Pope2000}. The coefficient can also be dynamically determined through scale-similarity assumption, leading to the dynamic Smagorinsky model (DSM) \cite{Germano1992}. However, the DSM requires explicit filtering of the flow field, and the type of filter must satisfy certain conditions \cite{Pope2000}. For wall-bounded turbulence, a well-known SGS model is the wall-adapting local eddy-viscosity (WALE) model \cite{Nicoud1999}, which is capable of recovering the near-wall scaling without any dynamic procedure, and free from the explicit filtering of the LES variables. Detailed derivations of these models can be well found in the literature and not reproduced \cite{Germano1992,Nicoud1999,Xie2020}.

%% Labels are used to cross-reference an item using \ref command.
% Section text. See Subsection \ref{subsec1}.

\section{\label{sec:HUFNO}The hybrid U-Net and Fourier neural operator (HUFNO)}

\begin{figure}\centering
\includegraphics[width=.75\textwidth]{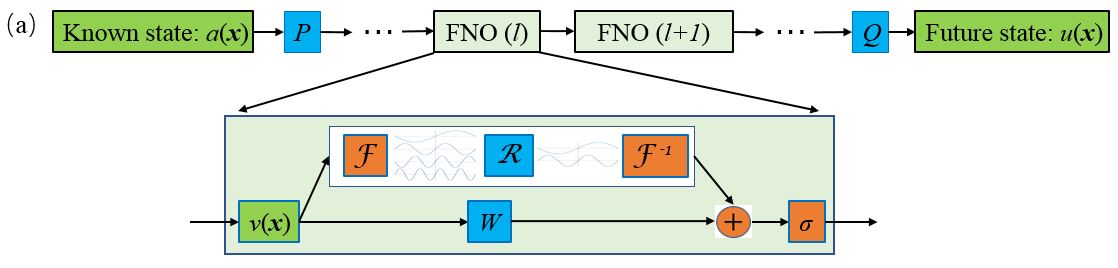}\hspace{-0.08in}
\includegraphics[width=.75\textwidth]{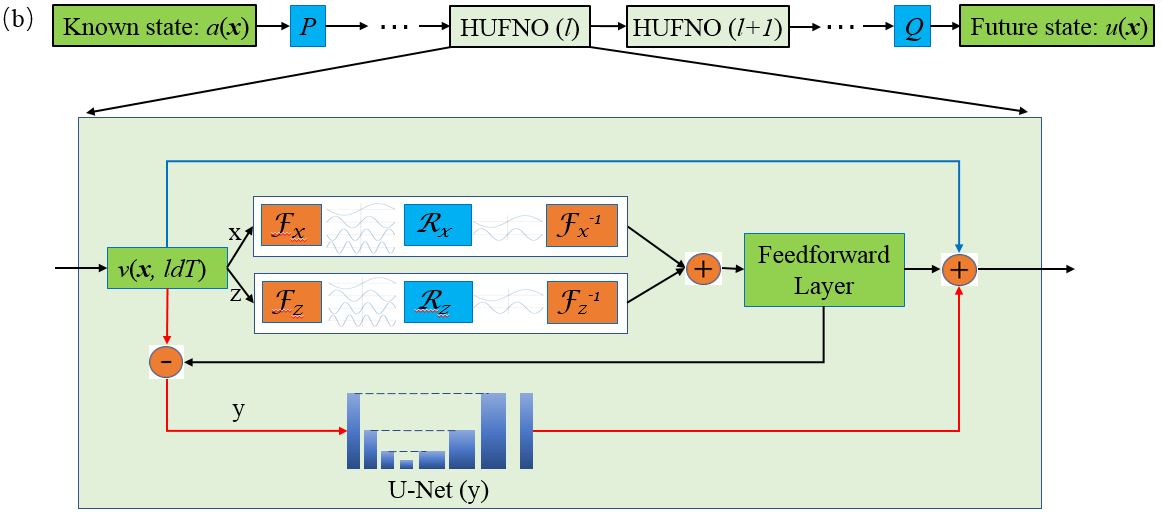}\vspace{-0.08in}
 \caption{The configurations of (a) Fourier neural operator (FNO); and (b) hybrid U-Net and Fourier neural operator (HUFNO).}\label{fig_config}
\end{figure}

While many NN-based methods focus on learning the nonlinear mappings of given input-output pairs in the physical domain, the FNO approach proposed by Li et al. learns the mappings of the high-dimensional features in the frequency domain \cite{Li2022}. Denote $D \subset \mathbb{R}^d$ as a bounded, open set and $\mathcal{A}=\mathcal{A}\left(D ; \mathbb{R}^{d_a}\right)$ and $\mathcal{U}=\mathcal{U}\left(D ; \mathbb{R}^{d_u}\right)$ as separable Banach spaces of function taking values in $\mathbb{R}^{d_a}$ and $\mathbb{R}^{d_u}$ respectively \cite{Beauzamy2011}. The mapping $\mathcal{A} \rightarrow \mathcal{U}$ can be approximated and parameterized in Fourier space as illustrated in Fig.~\ref{fig_config}a, where the input variables $a(\mathbf{x})$ denote the known states, and are projected to a higher dimensional representation $v(\mathbf{x})$ through a shallow fully connected neural network $P$. Subsequently, $v(\mathbf{x})$, taking values in $\mathbb{R}^{d_v}$, are updated between the Fourier layers according to
  \begin{equation}
  v_{m+1}(\mathbf{x})=\sigma\left(W v_m(\mathbf{x})+\left(\mathcal{K}(a ; \phi) v_m\right)(\mathbf{x})\right), \quad \forall \mathbf{x} \in D,  
  \label{v_update}
  \end{equation}
where $m$ denotes the $m$th Fourier layer, $\mathcal{K}: \mathcal{A} \times \Theta_{\mathcal{K}} \rightarrow \mathcal{L}\left(\mathcal{U}\left(D ; \mathbb{R}^{d_v}\right), \mathcal{U}\left(D ; \mathbb{R}^{d_v}\right)\right)$ maps to bounded linear operators on $\mathcal{U}\left(D ; \mathbb{R}^{d_v}\right)$ and is parameterized by $\phi \in \Theta_{\mathcal{K}}$, $W: \mathbb{R}^{d_v} \rightarrow \mathbb{R}^{d_v}$ is a linear transformation, and $\sigma: \mathbb{R} \rightarrow \mathbb{R}$ is the non-linear local activation function. Finally, the output function $u \in \mathcal{U}$ is obtained by $u(\mathbf{x})=$ $Q\left(v_m(\mathbf{x})\right)$ where $Q: \mathbb{R}^{d_v} \rightarrow \mathbb{R}^{d_u}$ is the projection of $v_m$, parameterized by a fully connected layer \cite{Li2020}.

Letting $\mathcal{F}$ and $\mathcal{F}^{-1}$ denote the forward and reverse Fourier transforms of a function $f: D \rightarrow \mathbb{R}^{d_v}$ respectively, and substituting the kernel integral operator in Eq.~\ref{v_update} with a convolution operator defined in Fourier space, one can write the Fourier integral operator as 
  \begin{equation}
  \left(\mathcal{K}(\phi) v_m\right)(\mathbf{x})=\mathcal{F}^{-1}\left(R_\phi \cdot\left(\mathcal{F} v_m\right)\right)(\mathbf{x}), \quad \forall \mathbf{x} \in D,
  \label{eq10}
  \end{equation}
with $R_\phi$ being the Fourier transform of a periodic function $\mathcal{K}: \bar{D} \rightarrow \mathbb{R}^{d_v \times d_v}$ parameterized by $\phi \in \Theta_{\mathcal{K}}$. The finite-dimensional parameterization is obtained by truncating the Fourier series at mode $k_{\max }$. $R_\phi$ is parameterized as complex-valued-tensor $({k_{\max } \times d_v \times d_v})$ containing a collection of truncated Fourier modes $R_\phi \in \mathbb{C}^{k_{\max } \times d_v \times d_v}$, where $\mathbb{C}$ is the complex space. 

In the classical FNO method, the Fourier transform (and its inverse) is performed in all three directions for 3D problems \cite{Li2022}, and this induces problems in the context of non-periodic boundary conditions even though some \emph{ad hoc} treatments would alleviate the problem \cite{Li2020,Wang2024}. Recently, Tran et al. has shown that the FNO can be factorized into different directions and performed separately \cite{Tran2021}. In this case, we can adopt the Fourier operations in the periodic directions while leaving the non-periodic direction(s) handled by the convolutional neural network (CNN) which does not require spatial periodicity. For the CNN, we choose the U-Net framework, whose symmetrical encoder-decoder structure and skip connections allow it to access low- and high-level features simultaneously \cite{Ronneberger2015}. This hybrid U-Net and FNO framework will be termed as HUFNO throughout the paper. In the U-Net of each HUFNO layer, the convolution kernel size = 3 and the stride = 2, leading to an automatic down-sampling with the encoding process \cite{Li2023}. The depth of the U-Net is 3, such that three consecutive encoding (convolution) layers are adopted followed by three consecutive decoding (deconvolution) layers with the classic skip connections. As we shall see, the HUFNO model outperforms both the original FNO and U-Net. The architecture of HUFNO is displayed in Fig.~\ref{fig_config}b. For HUFNO, we first define the operation $\mathcal{K}^{f}_{x,z}$as
\begin{equation}
% \begin{multline} % use for double columns, + \\ 
  \mathcal{K}^{f}_{x,z}(v_{l}(\mathbf{x}))=\mathcal{F}^{-1}_x(R_x\cdot \mathcal{F}_x (v_{l}(\mathbf{x}))) + \mathcal{F}^{-1}_z(R_z\cdot \mathcal{F}_z(v_{l}(\mathbf{x}))),
  \label{eq_FFFT_xz}
% \end{multline}
  \end{equation}
which contains the Fourier operations in the periodic $x$ and $z$ directions. $v_l(\mathbf{x})$ is the high-dimensional feature in the $l$th Fourier layer. Subsequently, the iterative updating procedure of $v(\mathbf{x})$ in HUFNO can be written as
\begin{equation}
  \begin{aligned}
    {v}_{l+1}(\mathbf{x}) & = \mathcal{U}_y({v}_{l}(\mathbf{x}) - \mathcal{L}(\mathcal{K}^f_{x,z}({v}_l(\mathbf{x})))) + \mathcal{L}(\mathcal{K}^{f}_{x,z}(v_{l}(\mathbf{x}))) + {v}_{l}(\mathbf{x}),
  \end{aligned}
  \label{eq_uffno}
\end{equation}
where the U-Net operation, denoted by $\mathcal{U}_y$, is performed in the $y$ direction, and $\mathcal{L}$ is a two-layer feed forward operation defined as \cite{Tran2021}
\begin{equation}
  \mathcal{L}(v_{l}(\mathbf{x})) = \sigma (W_2 \sigma (W_1 \mathcal{K}^f(v_{l}(\mathbf{x})) + b_1) + b_2),
  \label{eq_feedforward}
\end{equation}
where $\sigma$ is a nonlinear activation function. As evident in Eq.~\ref{eq_uffno}, the non-periodicity is handled by U-Net while the Fourier operations are only applied in the periodic directions. Meanwhile, a residual connection is incorporated, allowing an easier flow of gradients.

\section{\label{sec:post}Numerical tests in the turbulent flows over periodic hills}

\begin{figure}\centering
\includegraphics[width=.5\textwidth]{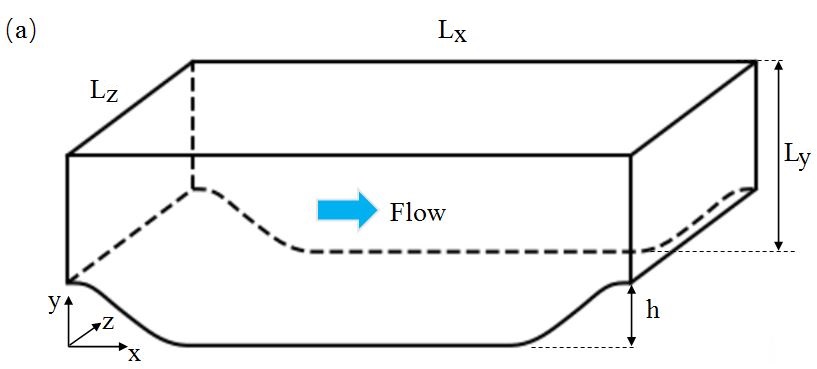}\hspace{-0.10in}
\includegraphics[width=.5\textwidth]{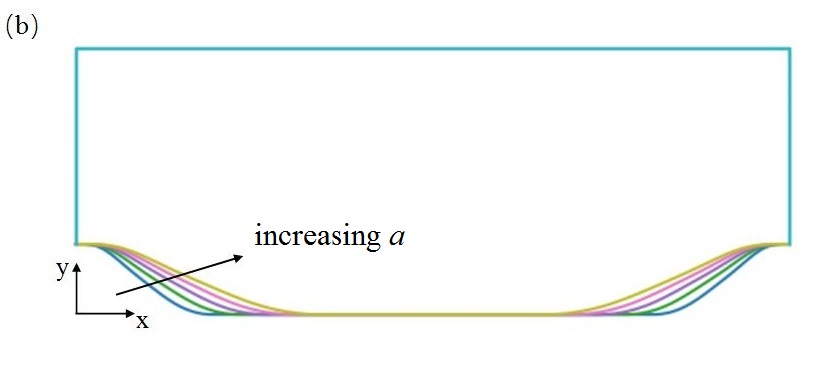}\vspace{-0.08in}
 \caption{The computational domain of periodic hill flow: (a) 3D view; (b) varying shapes in 2D view.}\label{fig_hill}
\end{figure}

In this section, we test the performance of the newly proposed HUFNO framework in the LES of turbulent flows over periodic hills. The computational domain of the periodic hill flow is displayed in Fig. \ref{fig_hill}a, where $h$ is the height of the hill. In this work, all length scales are normalized by the height of the hill. In this case, the streamwise, vertical and spanwise dimensions of the domain are $L_x=9$, $L_y=3.036$ and $L_z=4.5$ \cite{Breuer2009}. The velocities are normalized by the bulk velocity above the crest of the hill. To define the shape of hill, we introduce a new coordinate variable $x_L$, and let 
\begin{eqnarray}
  x_L &=& x,~~~~~~~ \text{for } x\le4.5,\nonumber\\
      &=& 9-x,~~ \text{for } x>4.5.
  \label{x_sym}
\end{eqnarray}

Then we define the local height of the hill $y^h(x_L)$ as
\begin{eqnarray}
 y^h(x_L) &=&min[1, 1 + 5.312(\frac{x_L}{a})^2 - 46.638(\frac{x_L}{a})^3], x\in[0,0.321),\nonumber\\
        &=&25.074 + 27.313(\frac{x_L}{a}) - 79.664(\frac{x_L}{a})^2 + 41.484(\frac{x_L}{a})^3, x\in[0.321,0.5),\nonumber\\
        &=&25.796 + 22.979(\frac{x_L}{a}) - 70.994(\frac{x_L}{a})^2 + 35.705(\frac{x_L}{a})^3, x\in[0.5,0.714),\nonumber\\
        &=&40.464 - 38.628(\frac{x_L}{a}) + 15.256(\frac{x_L}{a})^2 - 4.545(\frac{x_L}{a})^3, x\in[0.714,1.071),\nonumber\\
        &=&17.925 + 24.483(\frac{x_L}{a}) - 43.648(\frac{x_L}{a})^2 + 13.781(\frac{x_L}{a})^3, x\in[1.071,1.429),\nonumber\\
        &=&max[0, 56.390 - 56.295(\frac{x_L}{a}) + 12.896(\frac{x_L}{a})^2 + 0.587(\frac{x_L}{a})^3], x\in[1.429,1.929),\nonumber\\
        &=&0, x\in[1.929,4.5].\label{eq:hillshape}
\end{eqnarray}
Here a shape parameter $a$ is incorporated, such that the shape of hill can be varied as illustrated in Fig. \ref{fig_hill}b. A smaller shape factor $a$ indicating a steeper slope. With $a=1$, we recover the hill shape adopted by Breuer et al \cite{Breuer2009}. 

To obtain the training data as well as the benchmark data for LES, the DNS data is coarsened to LES grids. In the \emph{a posteriori} study, we compare the HUFNO-based solutions against the original FNO model, the U-Net, the benchmark DNS results as well as the traditional LES models. In the following sections, the DNS database and training data set are first introduced, followed by the \emph{a posteriori} studies in the LES.

%% Use \subsubsection, \paragraph, \subparagraph commands to 
%% start 3rd, 4th and 5th level sections.
%% Refer following link for more details.
%% https://en.wikibooks.org/wiki/LaTeX/Document_Structure#Sectioning_commands

\subsection{\label{subsec:training_config}The DNS database for model training}

The DNS database of turbulent flows over periodic hills is obtained using the open-source framework Xcompact3D, which is a high-order compact finite-difference CFD solver \cite{Laizet2009,Bartholomew2020}. An immersed boundary method (IBM) method is adopted to account for the presence of the solid hills. Periodic boundary conditions are adopted in the streamwise and spanwise directions. No-slip conditions are applied at the upper wall and the surface of the hill. Since the values of the convolution kernels in the U-Net are local and constant, uniform grids are adopted in the simulations for convenience. This also lowers the error when coarsening the DNS flow field to LES grids. Detailed DNS parameters are given in Table \ref{tab_dns}. In this study, three different Reynolds numbers are tested, namely $Re=700$, $1400$ and $5600$, such that the performance of the HUFNO model can be adequately assessed.
\begin{table*}
\begin{center}
\small
\begin{tabular*}{0.8\textwidth}{@{\extracolsep{\fill}}ccccccc}
\hline
$Re$ &Reso. &$\Delta x\times \Delta y\times \Delta z$ & $h$ &$ \Delta t$ \\ \hline
700 &$128\times129\times64$ &$0.070\times 0.024\times 0.070$ &1.0 &0.005  \\ 
1400 &$192\times193\times64$ &$0.047\times 0.016\times 0.070$ &1.0 &0.005 \\ 
5600 &$768\times385\times128$ &$0.012\times 0.004\times 0.035$ &1.0 &0.002 \\ \hline
\end{tabular*}
\normalsize
\caption{Parameters for the direction numerical simulation of turbulent flows over periodic hill.}\label{tab_dns}
\end{center}
\end{table*}

\subsection{\label{subsec:LES}Numerical experiments in the large-eddy simulation of periodic hill turbulence.}

\begin{figure}\centering
\includegraphics[width=.5\textwidth]{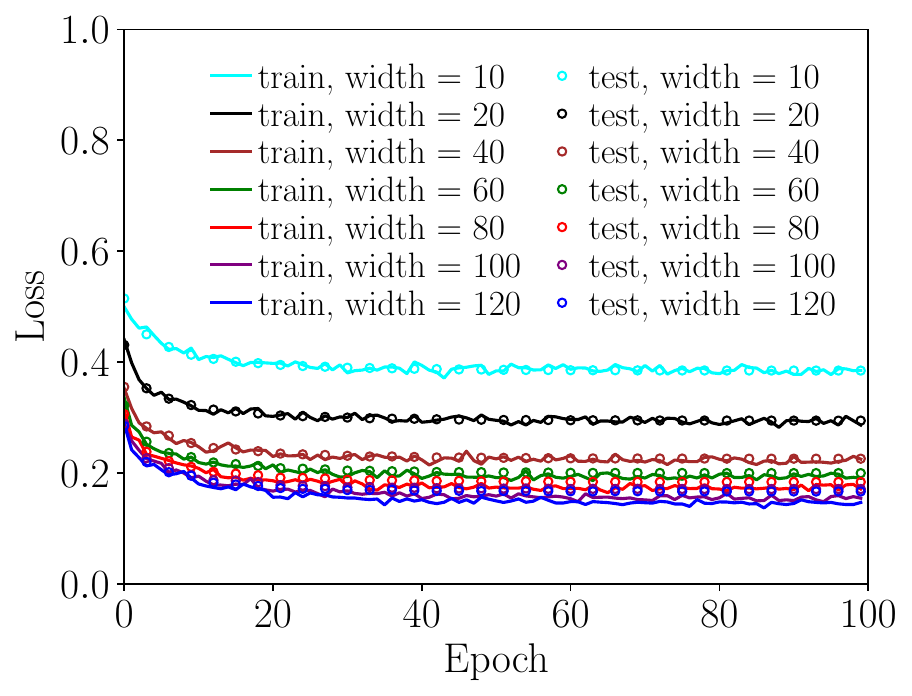}\hspace{-0.08in}
 \caption{The influence of the channel width in the high-dimensional space on the training and testing losses at $Re=700$.}\label{fig_width}
\end{figure}

In this study, the DNS data are coarsened to LES grids and taken as the benchmark for the LES. The computational grids in the LES are $32\times33\times16$, $48\times49\times16$ and $64\times65\times32$ for $Re=700$, $1400$ and $5600$, respectively. In the training datasets, the DNS data are extracted at every $\Delta T=200 \Delta t$ with $\Delta t$ being the DNS time step according to Table \ref{tab_dns}. Based on our tests, such configuration can give the best long-term performance. The known velocity fields and the shapes of the hill are taken as the input to the model. We take the known states of the previous five time nodes $[v(m-4), v(m-3), v(m-2), v(m-1), v(m)]$ as the model's inputs and the increment $v(m+1)-v(m)$ as the model's output. Five Fourier layers are adopted in this study. The Fourier modes are truncated at $k_{max}=5$, $5$ and $10$ for $Re=700$, $1400$ and $5600$, respectively, which are approximately 2/3 of the modes in the LES based the minimum grid numbers (i.e. in z direction). The channel width in the high-dimensional space after the projection of the inputs (cf. Fig. \ref{fig_config}) has a major influence on the performance of the model. The influence of the channel width on the training and testing losses is shown in Fig. \ref{fig_width} for $Re=700$. We observe that the value the testing loss saturates after width = 60, and we also note that the training and testing losses begin to separate at higher values of width (e.g. width = 120) indicating the potential risk of overfitting. Based on our tests, the influence of the Reynolds number on the sensitivity of the channel width is insignificant for the current problem. Hence, we set the channel width to 80. The Adam optimizer is adopted for optimization \cite{Kingma2014}, the initial learning rate is set to $10^{-3}$ which decays by half after every 10 iterations, and the ReLU function is chosen as the nonlinear activation function \cite{Glorot2011}.

In the \emph{a posteriori} tests, the flow fields are initialized using unseen turbulent conditions. Meanwhile, zero-velocity conditions are reinforced for the solid boundaries to ensure physical consistency. The prediction time step $\Delta T=200 \Delta t$, which is consistent with that in the training process. The statistics are calculated by time averaging the results. A total of 400 time steps are simulated, which allows the bulk flow to pass through the physical domain approximately 30 times. This ensures the convergence of the turbulence statistics for a stable simulation. In contrast, an unstable NN-based simulation can quickly diverge within this time range \cite{Yang2026,Wang2024}. In the following sections, we test the performance of HUFNO in four scenarios: section \ref{subsubsec:vary_ini} - at unseen initial conditions with fixed hill shapes, section \ref{subsubsec:vary_shape} - at unseen hill slopes, section \ref{subsubsec:unseen_Re} - at unseen Reynolds number, and section \ref{subsubsec:3d_shape} - in the case of three-dimensional hill (i.e. the shape of the hill varies in both the streamwise and spanwise directions).

\subsubsection{\label{subsubsec:vary_ini}Test at unseen initial conditions with fixed hill shapes}

\begin{figure}\centering
\includegraphics[width=.32\textwidth]{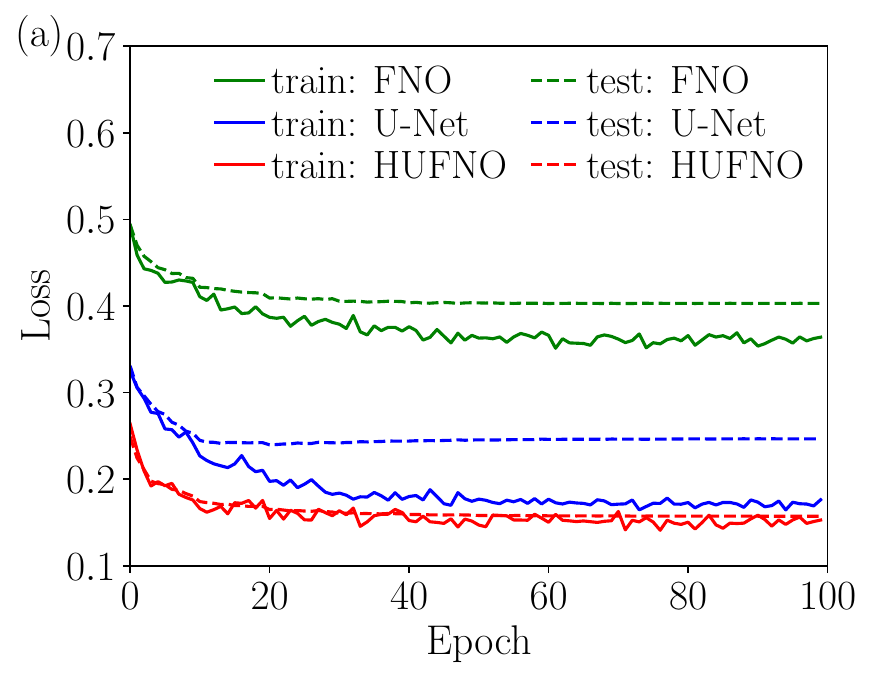}\hspace{-0.06in}
\includegraphics[width=.32\textwidth]{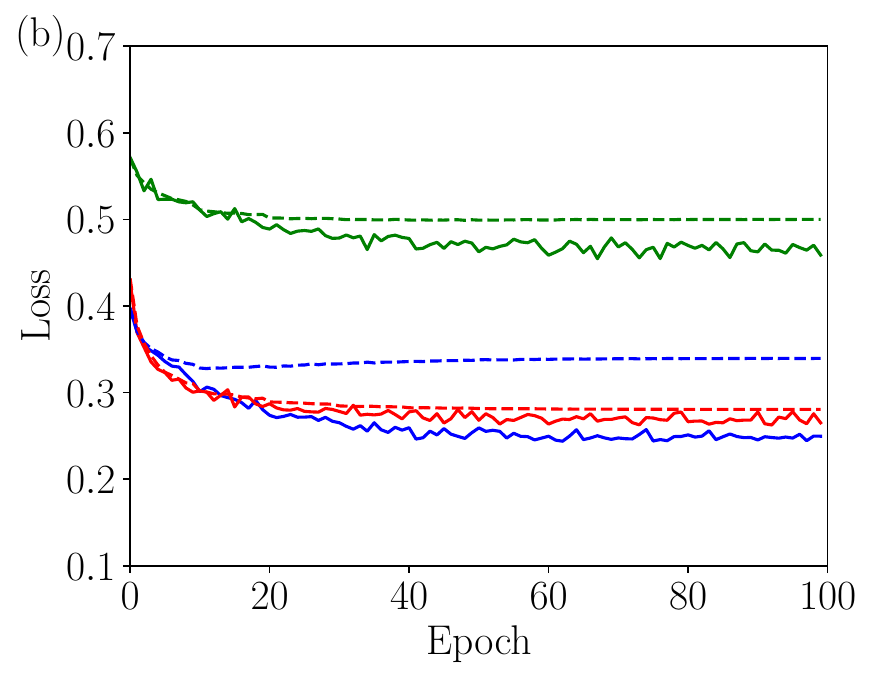}\hspace{-0.06in}
\includegraphics[width=.32\textwidth]{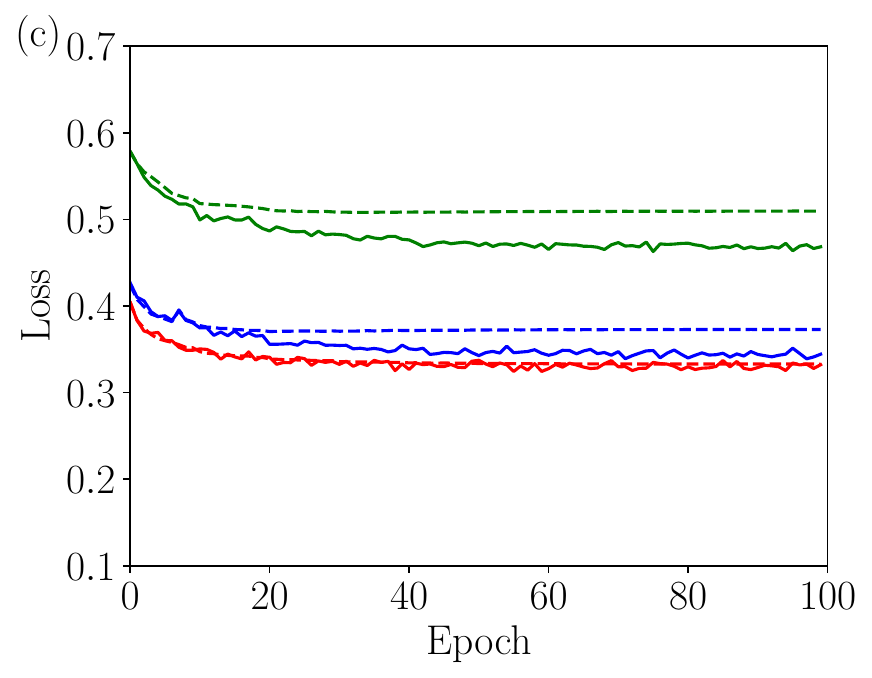}\vspace{-0.08in}
 \caption{The evolutions of the loss curves: (a) $Re=700$; (b) $Re=1400$; (c) $Re=5600$.}\label{fig_loss}
\end{figure}

In the simulation of period hill turbulence, The initial condition is constructed by superimposing random fluctuations onto a mean flow velocity profile, and this mean profile corresponds to a laminar channel flow profile over the crest of the hill, with the mean velocity being zero in the region beneath the hill crest. The random fluctuations are uniformly distributed over the range [-0.125, 0.125]. In this test, the hill shape is fixed at $a=1.0$, and twenty groups of DNS datasets are generated using different random initial fluctuations at each Reynolds number. The evolution history of the training and testing losses for the FNO, U-Net and HUFNO models are shown in Fig. \ref{fig_loss}. As can be seen, the HUFNO model has the lowest testing losses compared to the FNO and U-Net models, while the losses of FNO are the highest. We also note that the differences between the training and testing losses are the largest for the U-Net model, indicating potential overfitting. Meanwhile, for all three models, the losses increase with Reynolds number, indicating the increased prediction difficulty with higher Reynolds number.

\begin{figure}\centering
\includegraphics[width=.3\textwidth]{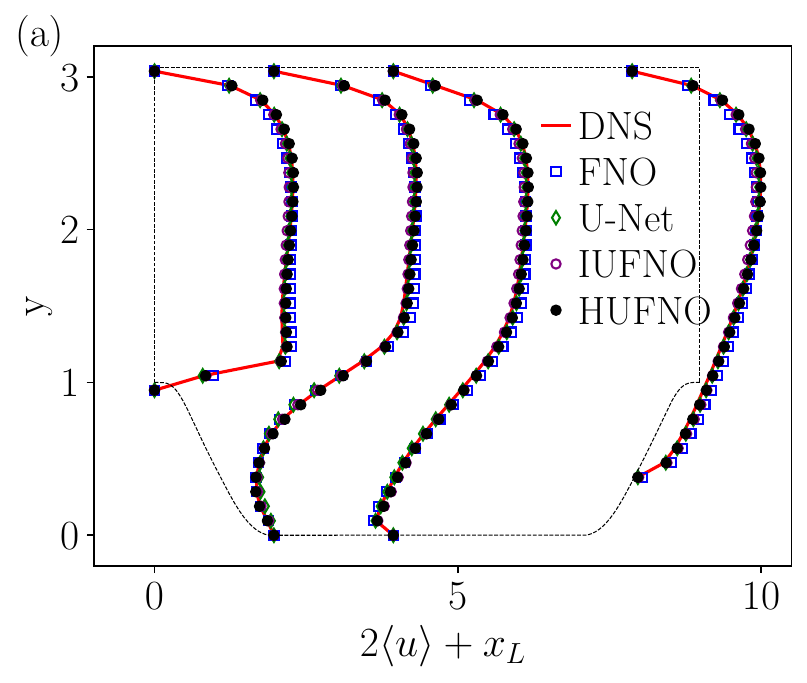}\hspace{-0.06in}
\includegraphics[width=.3\textwidth]{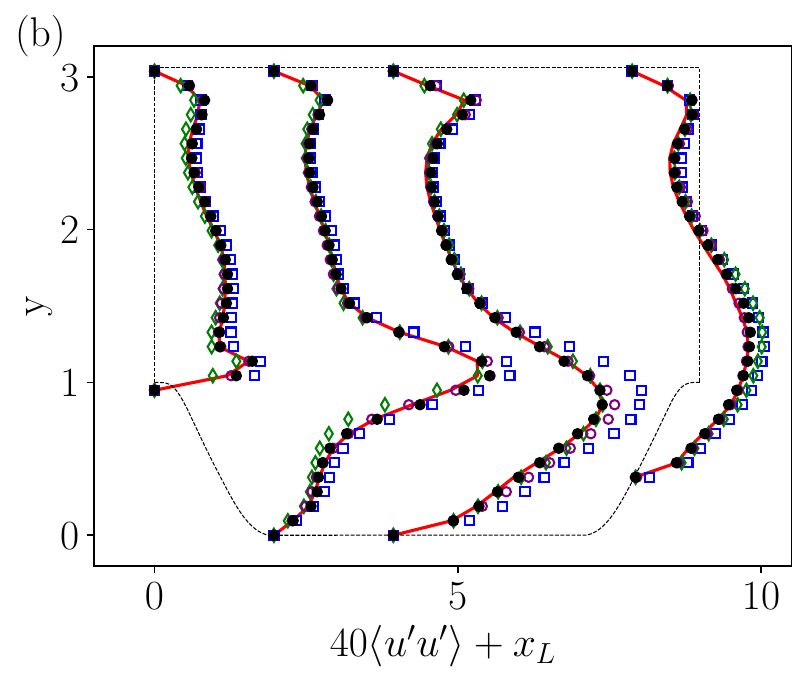}\hspace{-0.06in}
\includegraphics[width=.3\textwidth]{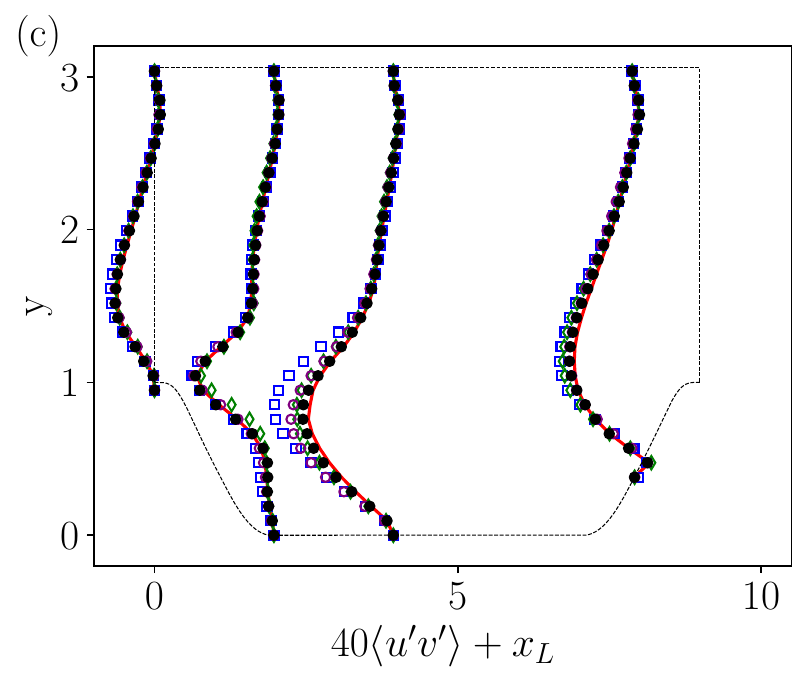}\vspace{-0.06in}

\includegraphics[width=.3\textwidth]{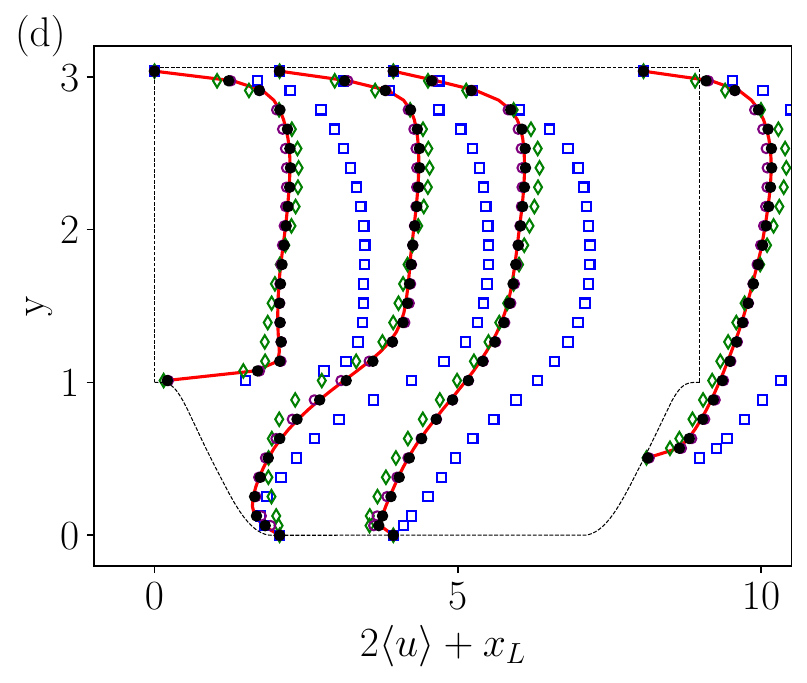}\hspace{-0.06in}
\includegraphics[width=.3\textwidth]{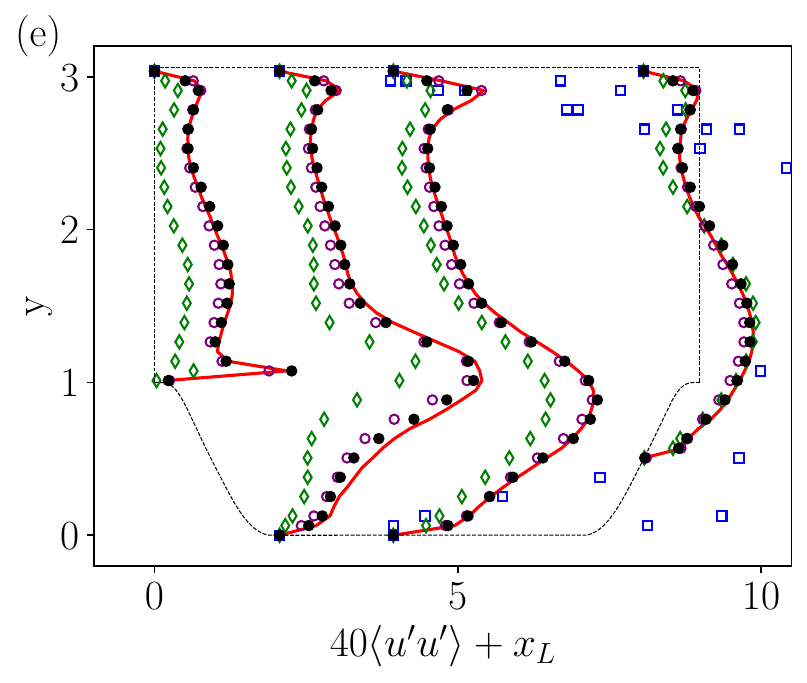}\hspace{-0.06in}
\includegraphics[width=.3\textwidth]{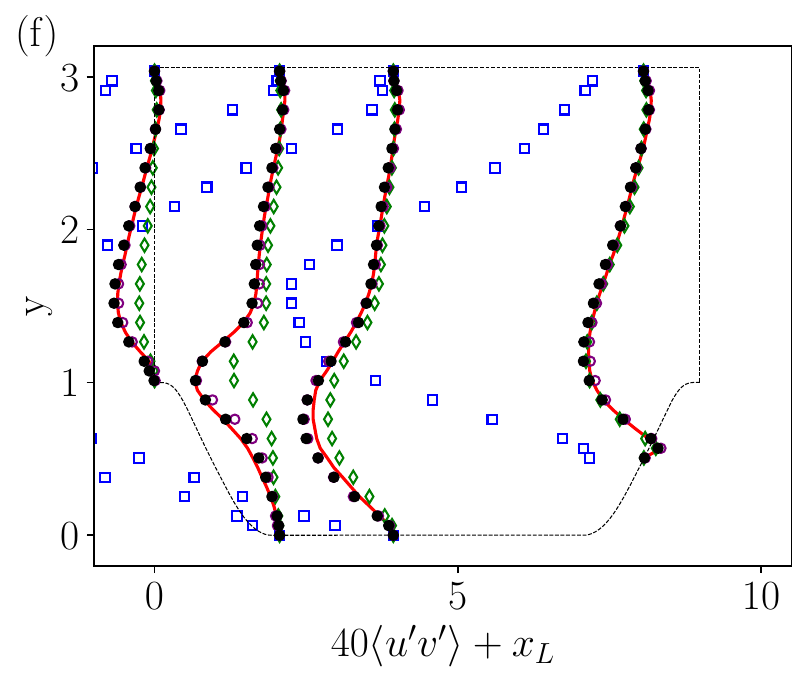}\vspace{-0.06in}

\includegraphics[width=.3\textwidth]{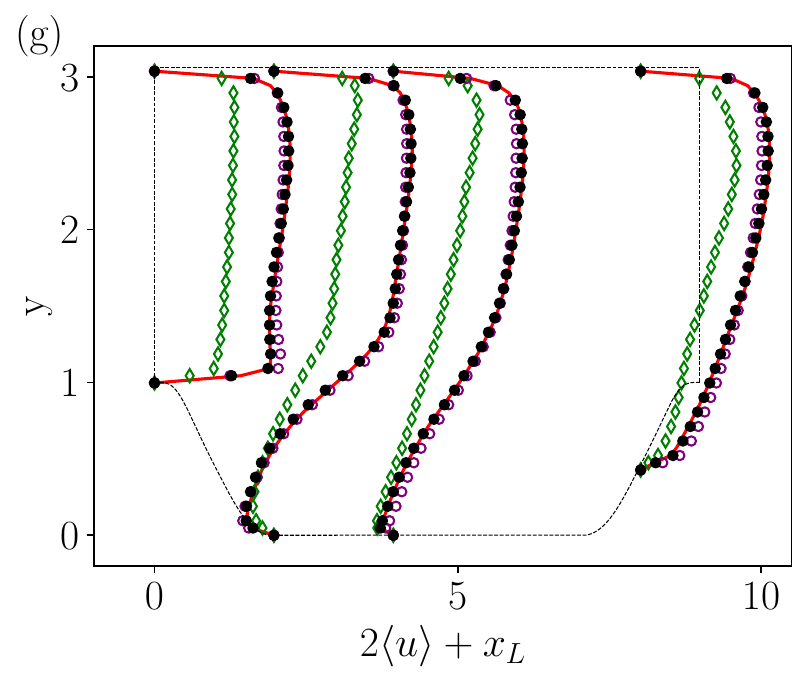}\hspace{-0.06in}
\includegraphics[width=.3\textwidth]{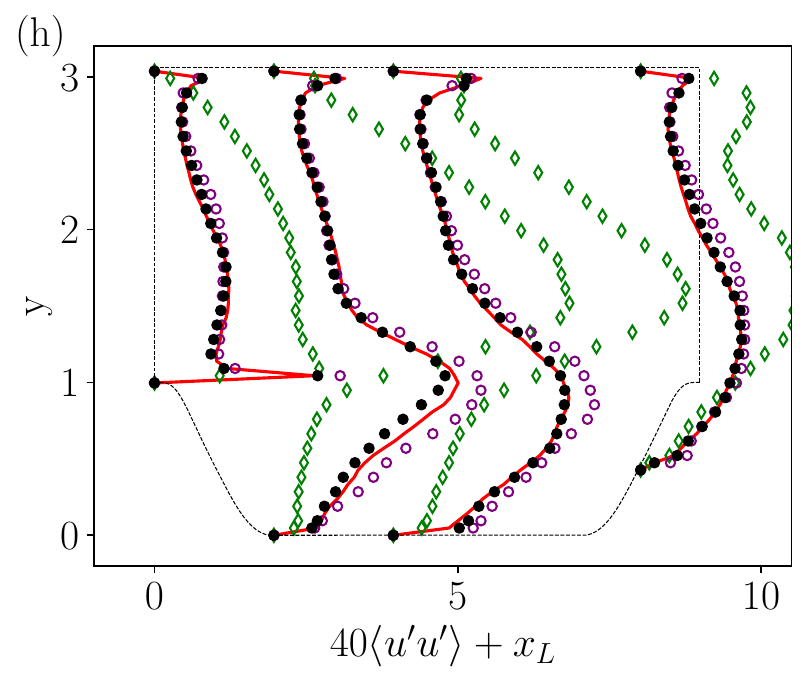}\hspace{-0.06in}
\includegraphics[width=.3\textwidth]{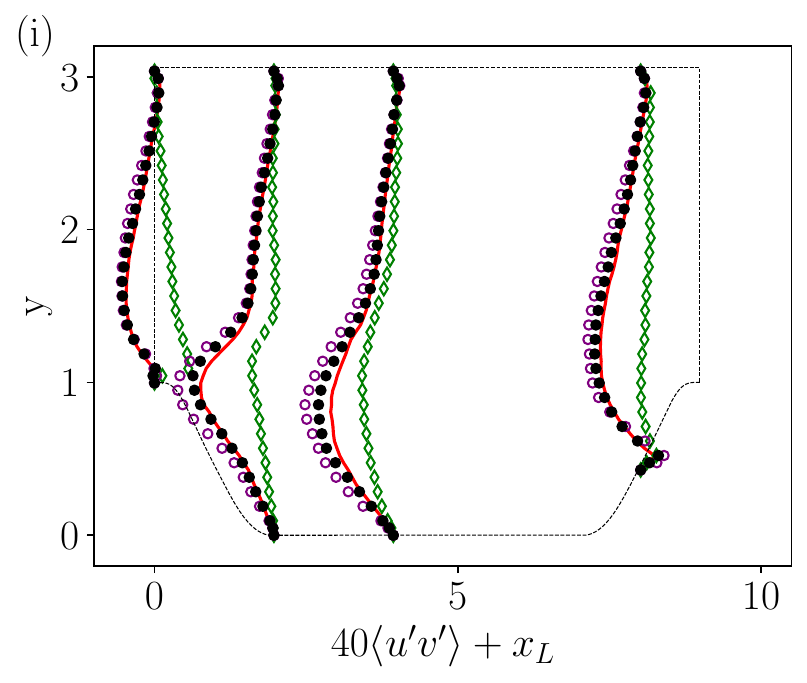}\vspace{-0.08in}
 \caption{The predicted turbulence statistics by different NN-based models: the first, second and third rows represent the results at $Re=700$, $1400$ and $5600$, respectively; the first, second and third columns represent $\langle u \rangle$, $\langle u'u' \rangle$ and $\langle u'v' \rangle$, respectively.}\label{fig_comparefno}
\end{figure}

In the \emph{a posteriori} tests, initial conditions different from the training sets are used to initialize the flow field. In Fig. \ref{fig_comparefno}, we compare the predictions of the turbulence statistics including the mean streamwise velocities $\langle u \rangle$, the normal Reynolds stresses $\langle u'u' \rangle$ and the shear Reynolds stress $\langle u'v' \rangle$ at different Reynolds numbers. Clearly, the HUFNO predictions have the closest agreements with the DNS benchmarks, underscoring its great efficacy. The FNO and U-Net have some small discrepancies at $Re=700$, while strongly deviate from the DNS results at high Reynolds numbers. Also shown in Fig. \ref{fig_comparefno} are the results predicted by our previously proposed IUFNO model. As can be seen, while the IUFNO outperforms the FNO and U-Net models, it is not as accurate as the HUFNO especially at higher Reynolds numbers. We note that, in the IUFNO model, the Fourier operations are performed in both the periodic and non-periodic directions, and an additional U-Net is incorporated and expected to correct the error associated with the non-periodicity. In comparison, the HUFNO model completely avoids the Fourier operations in the non-periodic direction, which not only reduces the number of NN parameters in the model but also makes the model more robust. Meanwhile, the FNO model diverges at $Re=5600$, and hence the corresponding results are not shown in the figure. As the HUFNO model gives the best performance, only the results of HUFNO will be presented and compared against the traditional LES models in the rest of the paper.

\begin{figure}\centering

\includegraphics[width=.3\textwidth]{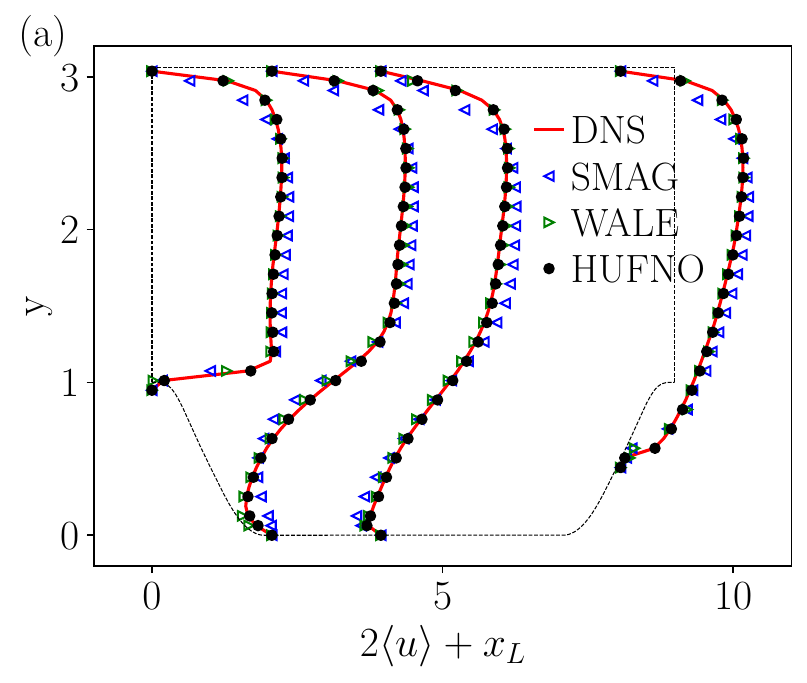}\hspace{-0.06in}
\includegraphics[width=.3\textwidth]{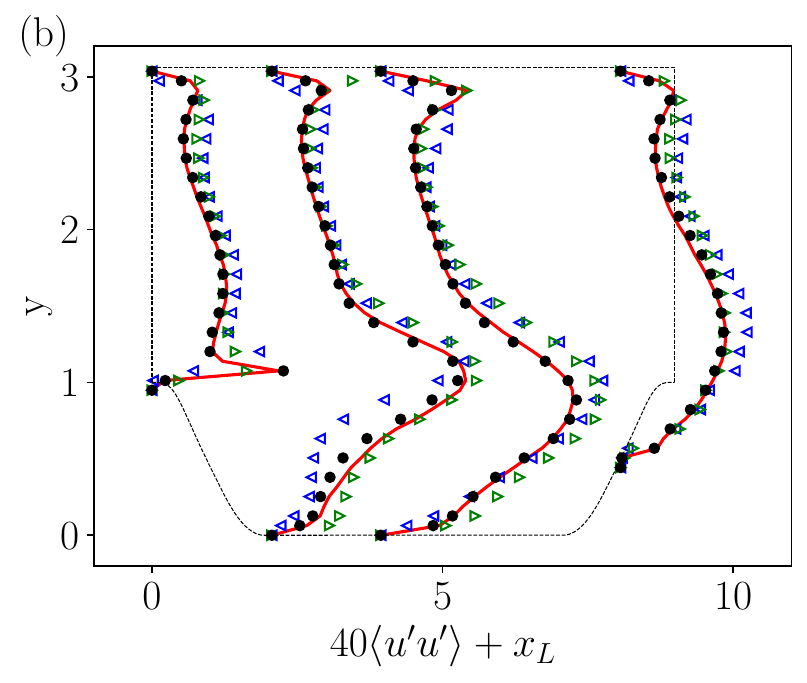}\hspace{-0.06in}
\includegraphics[width=.3\textwidth]{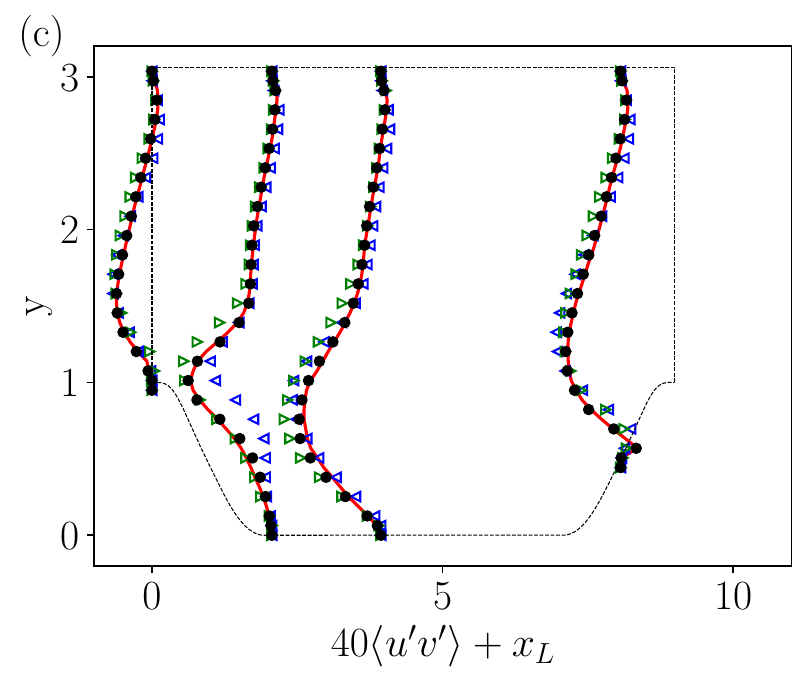}\vspace{-0.08in}

 \caption{The predicted turbulence statistics at $Re=1400$: (a) mean velocity $\langle u \rangle$; (b) normal Reynolds stress $\langle u'u' \rangle$; (c) shear Reynolds stress $\langle u'v' \rangle$.}\label{fig_compareles_velstat}
\end{figure}

The comparisons of the turbulence statistics between HUFNO model and the traditional LES are shown in Fig. \ref{fig_compareles_velstat} for $Re=1400$. Here the SMAG and WALE models are adopted in the traditional LES. As the figure depicts, the HUFNO model performs reasonably better compared to the SMAG and WALE models. Meanwhile, we observe that the WALE model can also predict the mean velocity very well, demonstrating its advantage over the SMAG model. However, in the predictions of the Reynolds stresses, both the SMAG and WALE models have noticeable deviations from the DNS results.

\begin{figure}\centering
\includegraphics[height=.24\textwidth]{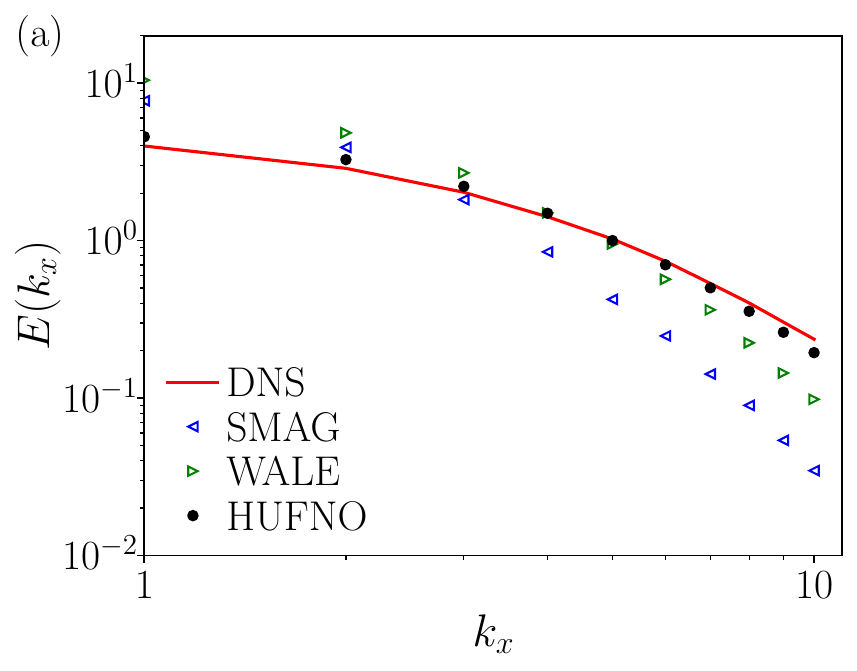}\hspace{-0.06in}
\includegraphics[height=.24\textwidth]{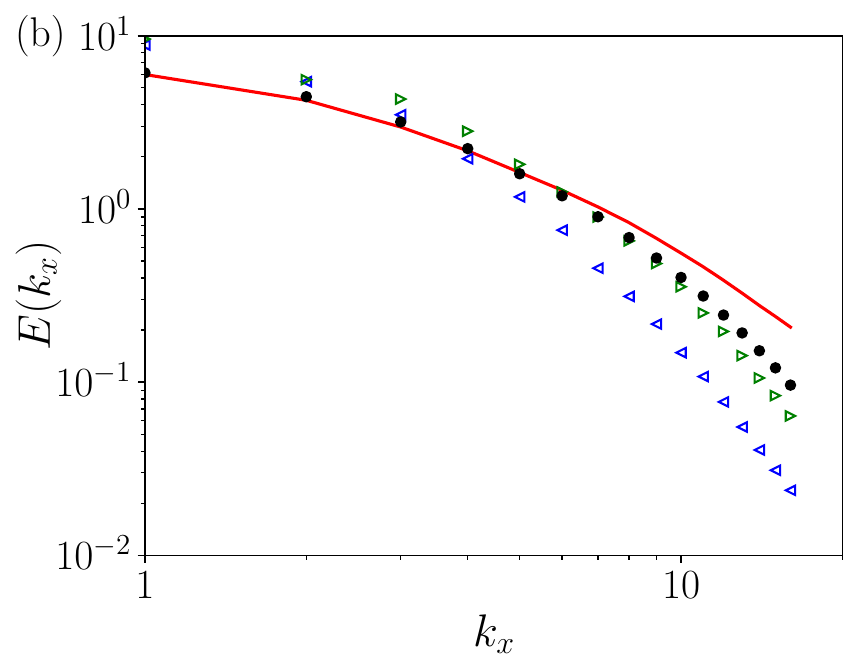}\hspace{-0.06in}
\includegraphics[height=.24\textwidth]{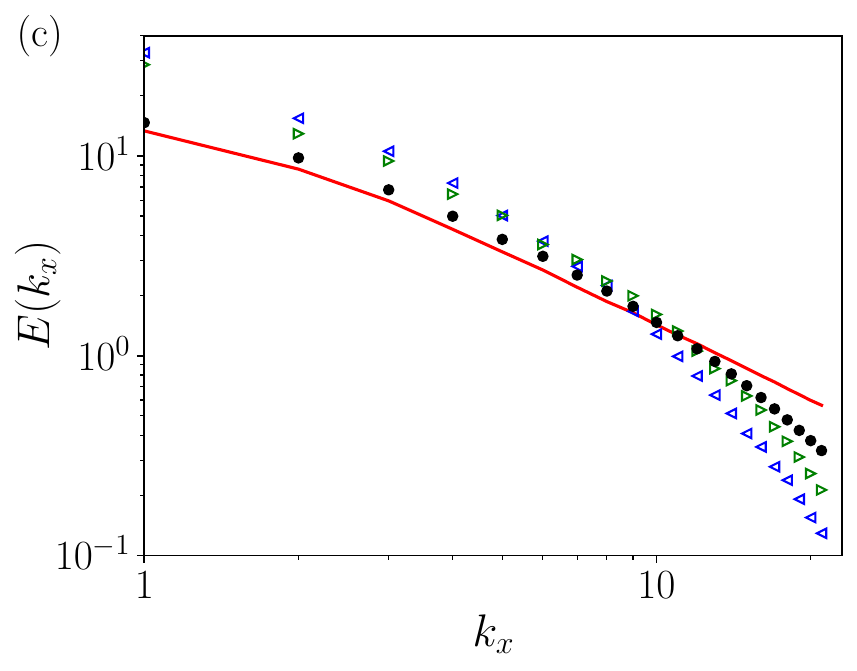}\vspace{-0.08in}

\caption{The predicted streamwise energy spectra: (a) $Re=700$; (b) $Re=1400$; (c) $Re=5600$.}\label{fig_Ek}
\end{figure}

To examine the energy distribution at different length scales, we calculate the streamwise kinetic energy spectrum in the LES. The results are shown in Figs. \ref{fig_Ek}a, \ref{fig_Ek}b and \ref{fig_Ek}c for $Re=700$, $1400$ and $5600$, respectively. At $Re=700$, the predicted spectrum by the HUFNO model agrees well with the DNS result. Both the SMAG and WALE models overestimate the large-scale energy while underestimate the small-scale energy. At $Re=1400$ and $5600$, the agreements between HUFNO and DNS are slightly worse compared to the case of $Re=700$, indicating increased prediction difficulty at higher Reynolds numbers. Nevertheless, the HUFNO model still tangibly outperforms the traditional SMAG and WALE models.

\begin{figure}\centering
\includegraphics[height=.245\textwidth]{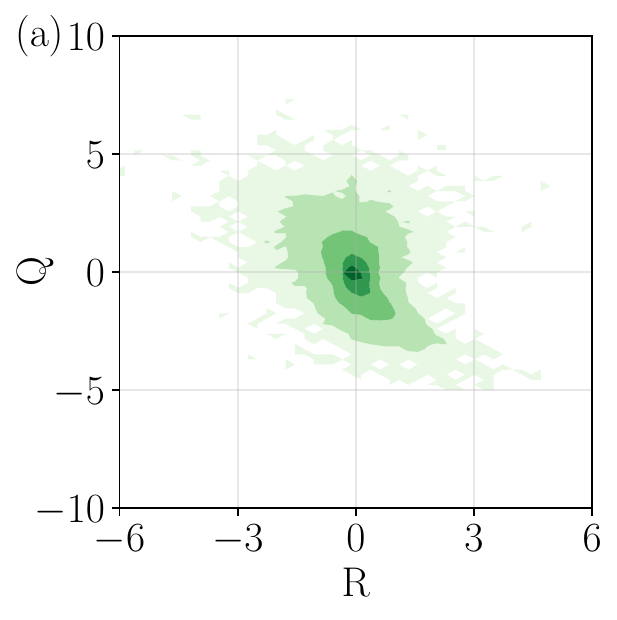}\hspace{-0.08in}
\includegraphics[height=.245\textwidth]{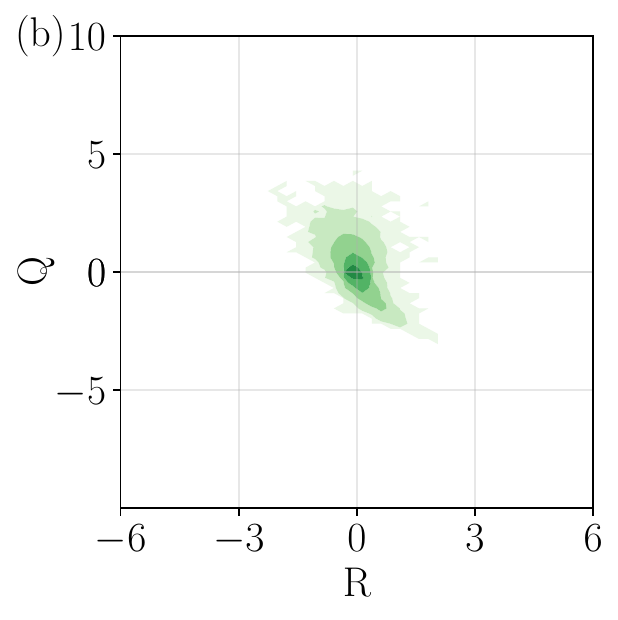}\hspace{-0.08in}
\includegraphics[height=.245\textwidth]{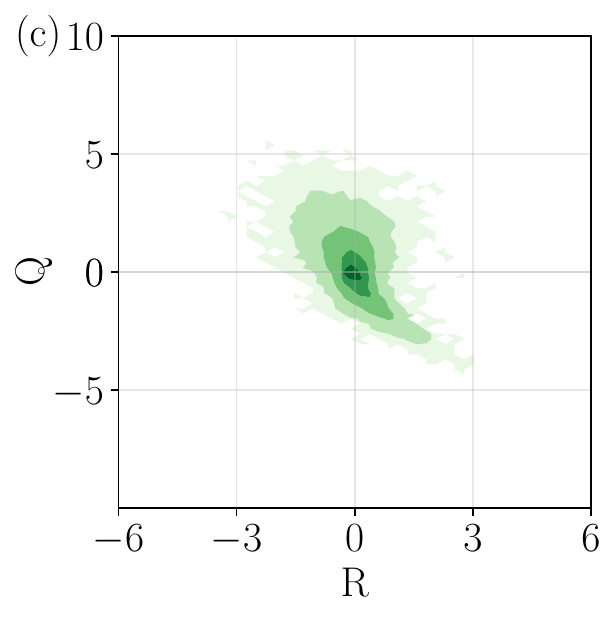}\hspace{-0.08in}
\includegraphics[height=.245\textwidth]{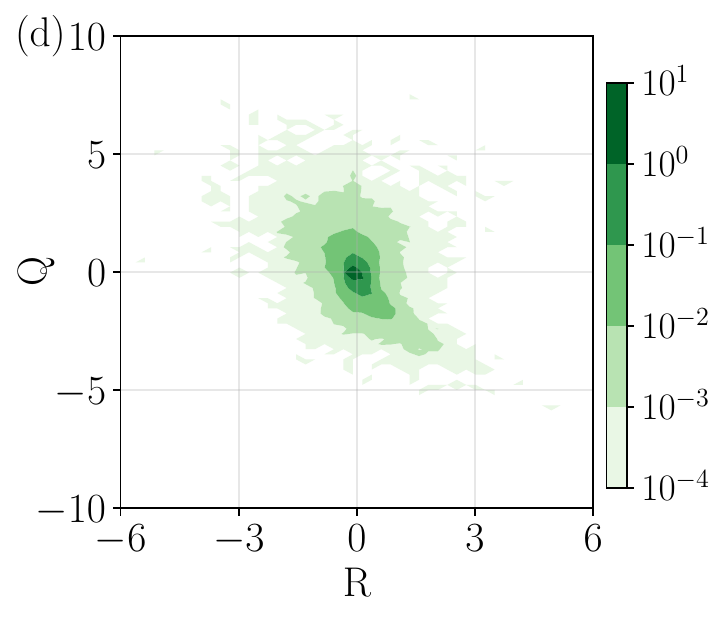}\vspace{-0.08in}
 \caption{The predicted $Q-R$ joint PDFs at $Re=700$: (a) DNS; (b) SMAG; (c) WALE; (d) HUFNO.}\label{fig_compareles_QR_test_20}
\end{figure}

The above results have demonstrated the capability of the HUFNO model in terms of accurately predicting the turbulence statistics and recovering the multiscale energy distributions. To further quantify the model's ability in learning physically meaningful turbulent flow features, we analyze the $Q-R$ invariant map\cite{Chong1990,Blackburn1996,Wang2020a}, which is widely used to characterize turbulent flow structures and identify different flow topologies. The $Q$ and $R$ are the second and third invariants of the velocity gradient tensor (VGT), which are defined as
 \begin{equation}
Q = \frac{1}{2} \left( |\Omega|^2 - |S|^2 \right),\ \   R = -\det(\nabla\mathbf{u}),
\end{equation}\label{QR_invariant}
respectively, where $|\Omega|^2 = \Omega_{ij}\Omega_{ij}$ and $|S|^2 = S_{ij}S_{ij}$ are the squared magnitudes of the rotation-rate tensor and the strain-rate tensor, respectively. For demonstration, we show the $Q-R$ joint PDF in Fig. \ref{fig_compareles_QR_test_20} for $Re=700$. The joint PDF takes a tear drop shape with a preference for the second and fourth quadrants, in consistency with existing findings for different type of flows \cite{Blackburn1996,Wang2020a}. Meanwhile, we observe that the HUFNO predictions most closely match the DNS benchmark. The prediction based on the WALE model is slightly less accurate than that of HUFNO, whereas the SMAG model yields the least accurate results, significantly underestimating extremes in both $Q$ and $R$ values. These results demonstrate HUFNO’s capability to accurately capture key physical characteristics of turbulent flows.

\begin{figure}\centering
\includegraphics[height=.24\textwidth]{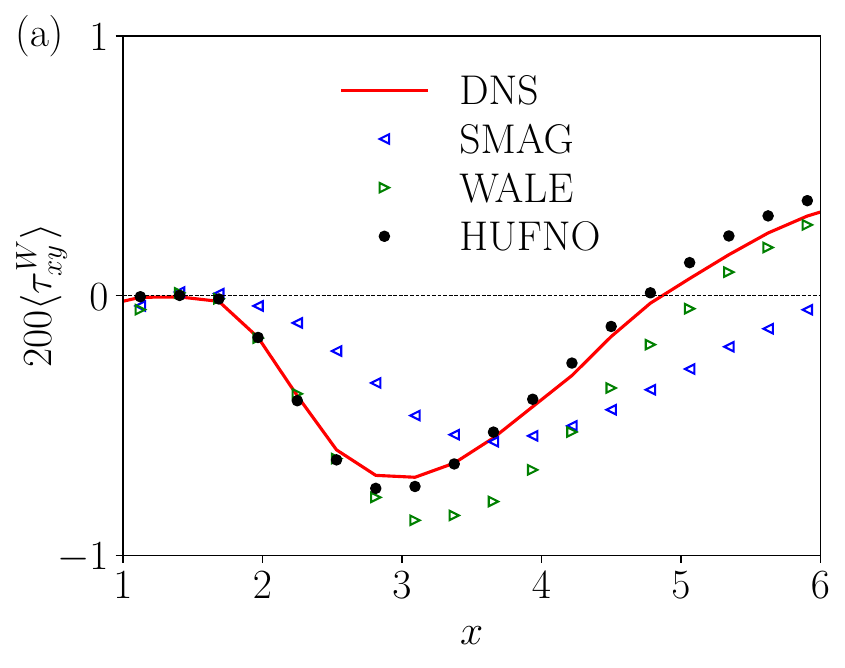}\hspace{-0.06in}
\includegraphics[height=.24\textwidth]{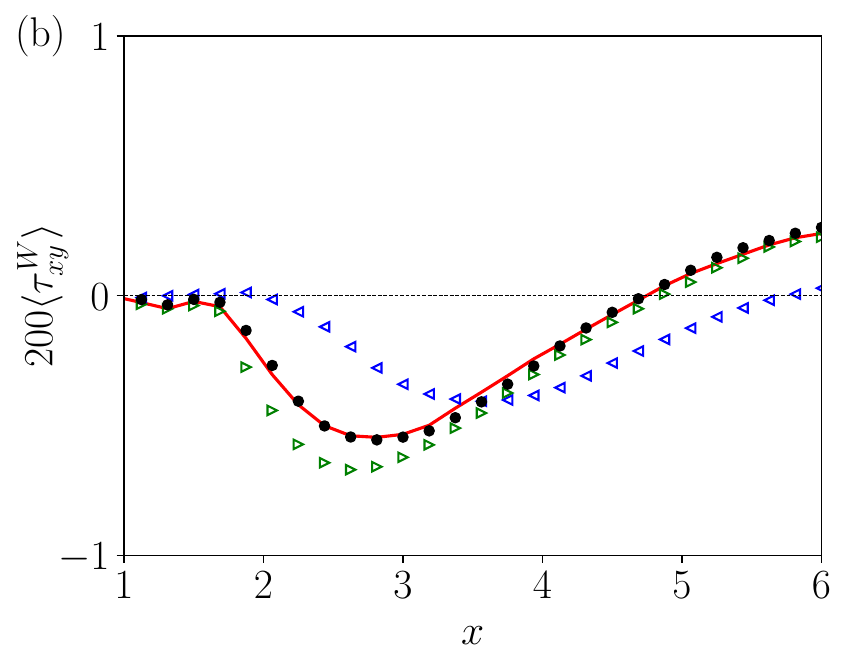}\hspace{-0.06in}
\includegraphics[height=.24\textwidth]{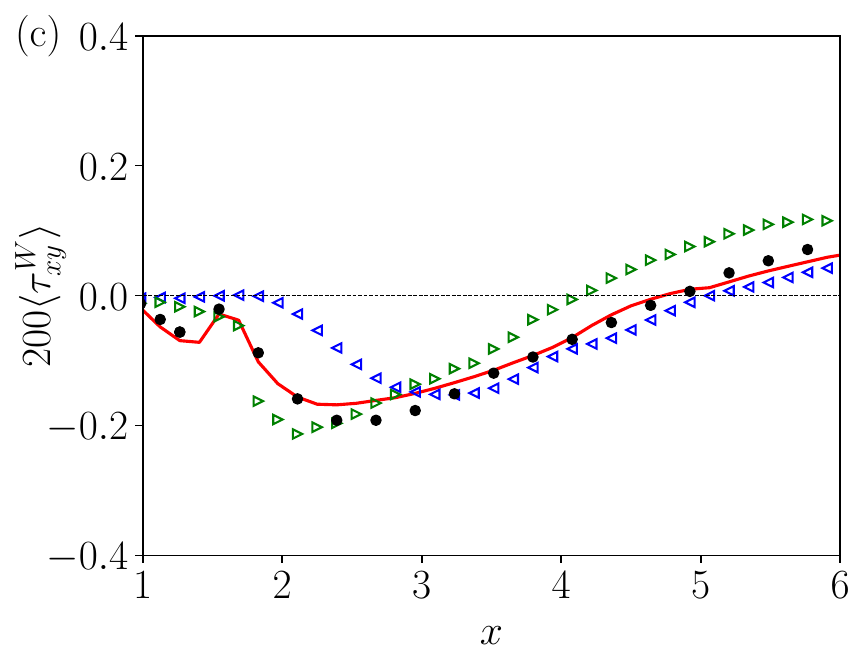}\vspace{-0.08in}
 \caption{The wall-shear stress in the flow separation region: (a) $Re=700$; (b) $Re=1400$; (c) $Re=5600$.}\label{fig_compareles_tauw}
\end{figure}

To assess the quality of predictions near the wall, we examine the dimensionless viscous wall-shear stress $\langle\tau^W_{xy}\rangle=\langle \partial u/\partial y \rangle/Re$ in the flow separation region. As shown in Fig. \ref{fig_compareles_tauw}, the intersection of the wall-shear stress and zero indicates the extents of the flow separation region. Compared to the traditional LES models, the HUFNO can more accurately predict the distribution of wall-shear stress on the hill surface. Meanwhile, the WALE model tangibly outperforms the SMAG model due to its physically consistent treatment of near-wall eddy viscosity \cite{Nicoud1999}.

\begin{figure}\centering
\includegraphics[width=.85\textwidth]{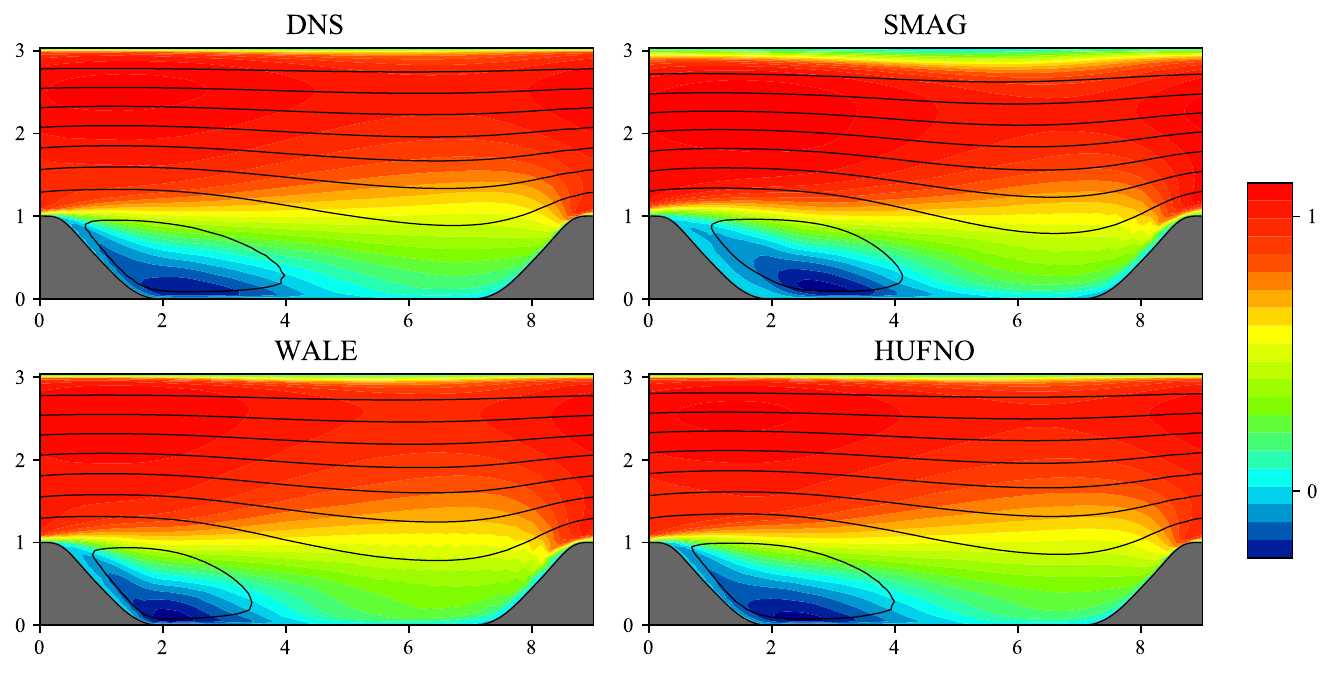}\vspace{-0.08in}
 \caption{The streamlines of the periodic hill flow at $Re=5600$. The domain is colored by the streamwise velocity.}\label{fig_compareles_streamline_re5600}
\end{figure}

To visualize the predicted flow structures of periodic hill turbulence, we present the flow streamline contours in Fig. \ref{fig_compareles_streamline_re5600} for $Re=5600$. The domain is colored by the streamwise velocity. The separation regions and the re-circulation streamlines can be clearly identified. As observed, the HUFNO model gives the best predictions for the flow separation structure, demonstrating its capability for the modeling of strongly separated flows.

In Table \ref{tab_cost}, we list the computational costs (in seconds) for the HUFNO model and traditional LES models. The HUFNO-based simulations are conducted on the NVIDIA A100 GPU, for which the CPU configuration is AMD EPYC 7763 @2.45GHz. The traditional LES simulations using the SMAG and WALE models are performed on the CPU of Intel Xeon Gold 6148 @2.40 GHz. As evidenced in the table, the computational efficiency of the HUFNO model is considerably higher than those of the traditional LES models, considering that the costs for the traditional LES models are not multiplied by the number of computational cores (given in the parentheses).

\begin{table*}
\begin{center}
\small
\begin{tabular*}{0.8\textwidth}{@{\extracolsep{\fill}}ccccccc}
\hline
$Re$ &SMAG &WALE &HUFNO \\ \hline
700 &66.23s ($\times$64 cores) &74.77s ($\times$64 cores) &2.02s \\ 
1400 &232.14s ($\times$64 cores) &249.71s ($\times$64 cores) &2.23s \\ 
5600 &270.00s ($\times$64 cores) &391.43 ($\times$64 cores) &6.33s \\ \hline
\end{tabular*}
\normalsize
\caption{Computational costs of different LES models per 10000 DNS time step.}\label{tab_cost}
\end{center}
\end{table*}

\subsubsection{\label{subsubsec:vary_shape}Test at unseen slopes of the hill}
%% Inline mathematics is tagged between $ symbols.

\begin{figure}\centering
\includegraphics[width=.3\textwidth]{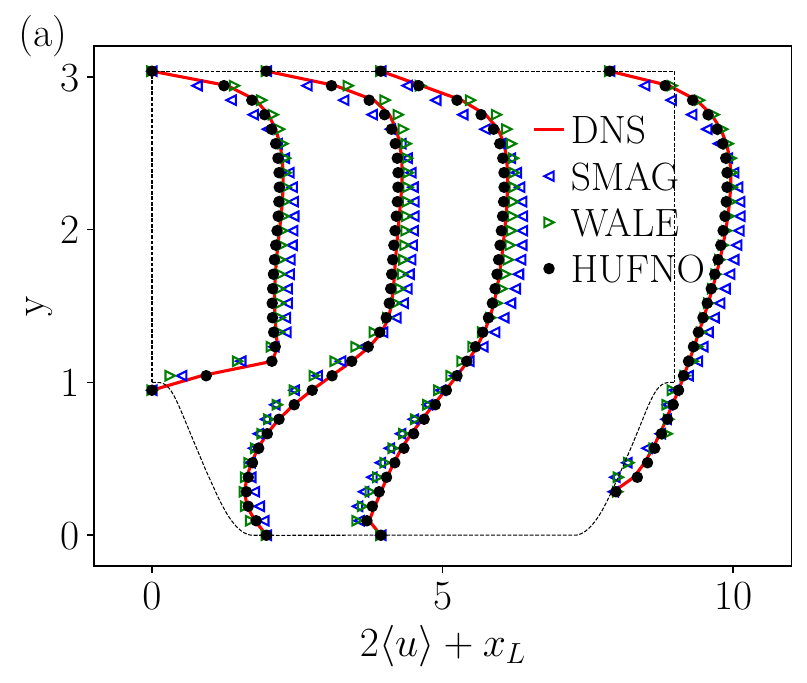}\hspace{-0.06in}
\includegraphics[width=.3\textwidth]{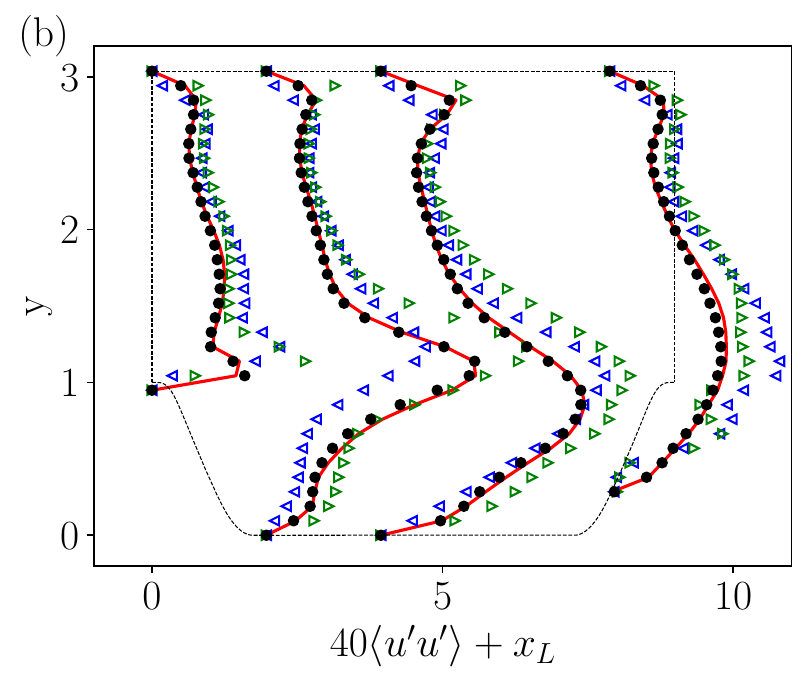}\hspace{-0.06in}
\includegraphics[width=.3\textwidth]{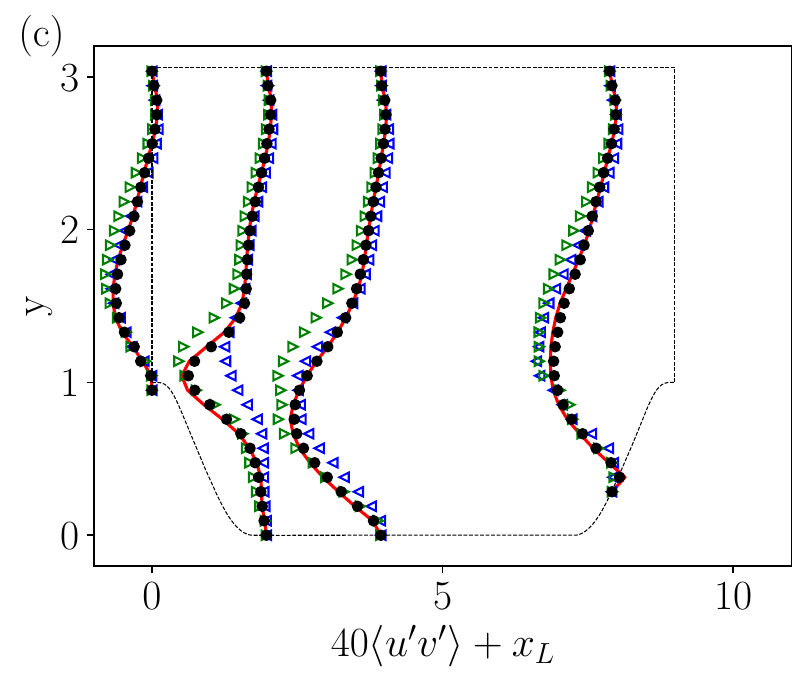}\vspace{-0.06in}

\includegraphics[width=.3\textwidth]{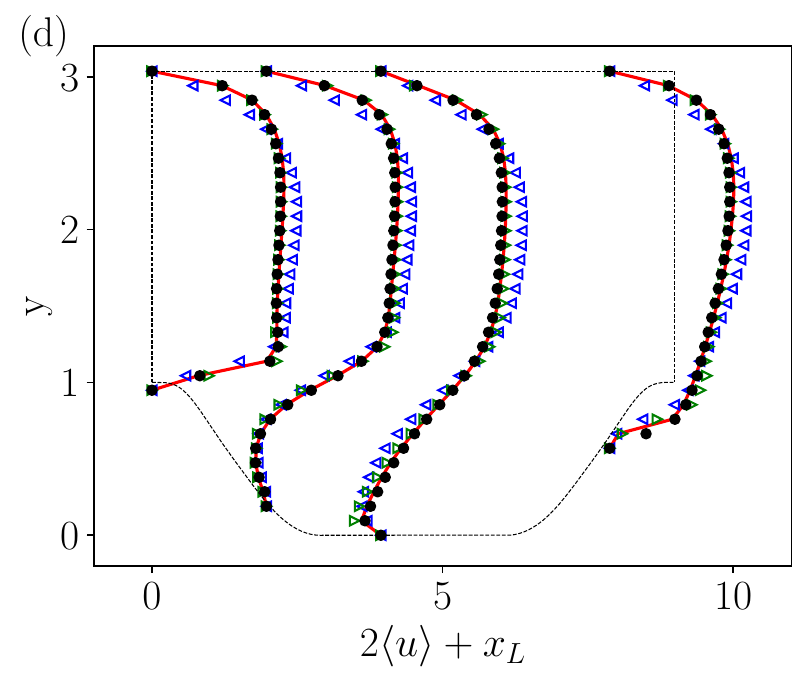}\hspace{-0.06in}
\includegraphics[width=.3\textwidth]{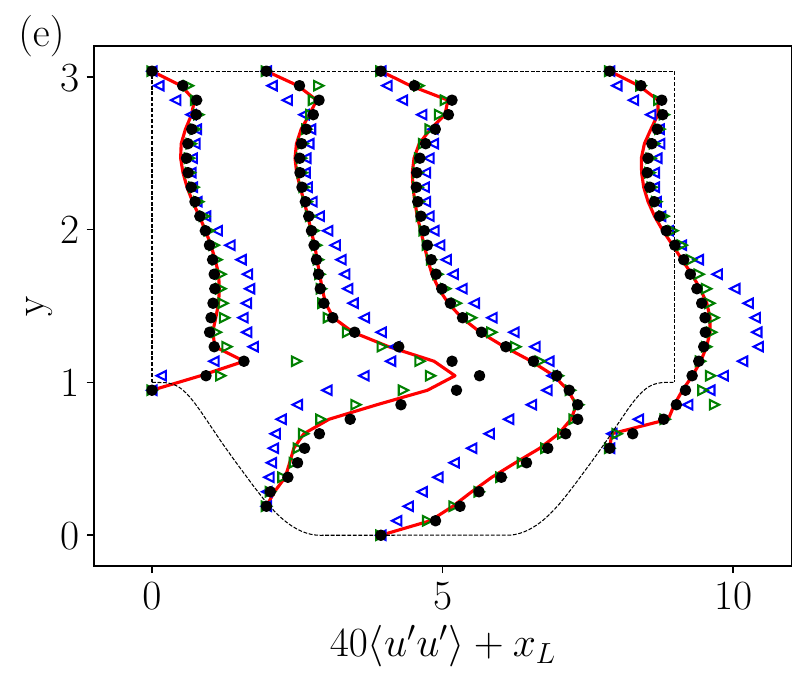}\hspace{-0.06in}
\includegraphics[width=.3\textwidth]{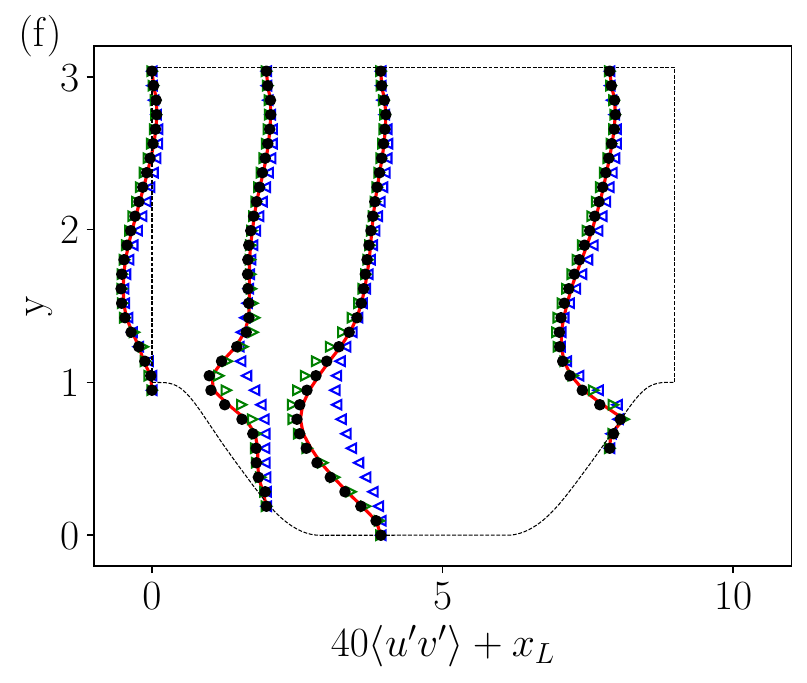}\vspace{-0.06in}

\includegraphics[width=.3\textwidth]{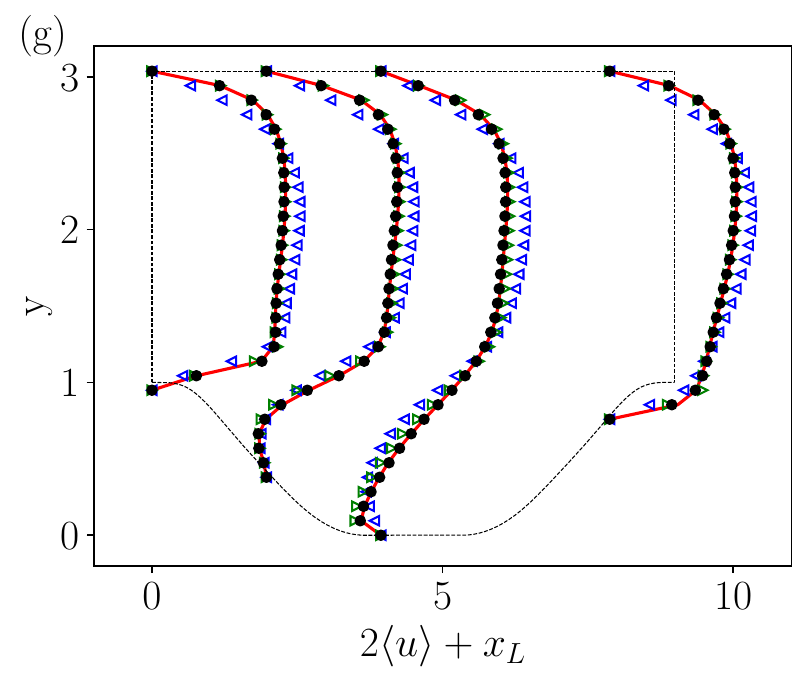}\hspace{-0.06in}
\includegraphics[width=.3\textwidth]{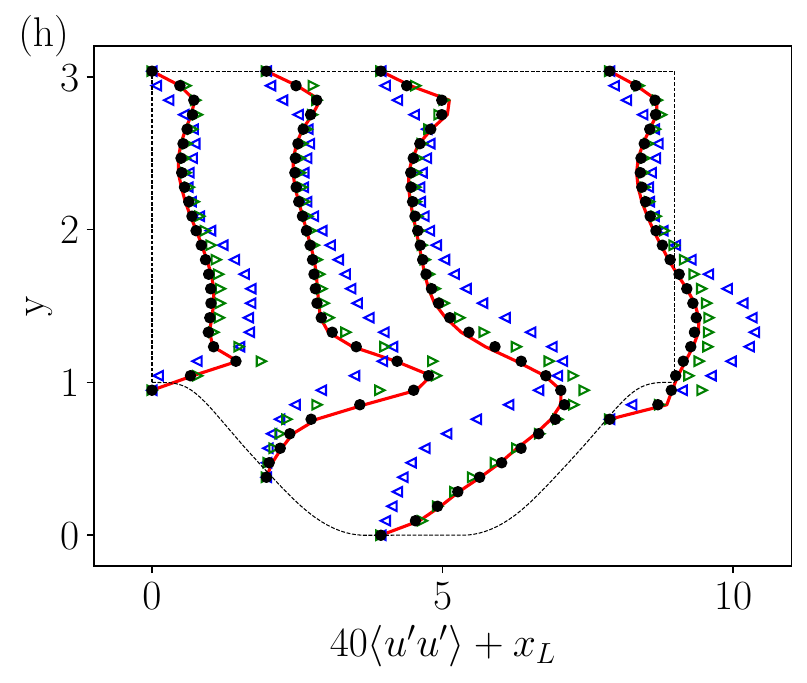}\hspace{-0.06in}
\includegraphics[width=.3\textwidth]{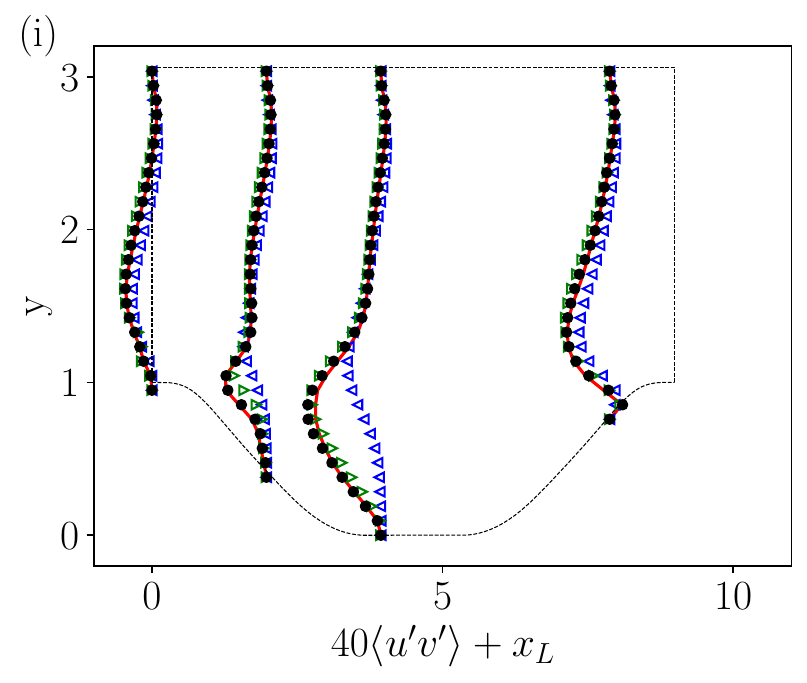}\vspace{-0.08in}

 \caption{The predicted turbulence statistics with varying hill shapes: the first, second and third rows represent the results for shape factors $a=0.9$, $1.5$ and $1.9$, respectively; the first, second and third columns represent the results for $\langle u \rangle$, $\langle u'u' \rangle$ and $\langle u'v' \rangle$, respectively.}\label{fig_compareles_velstat_varyshape}
\end{figure}

In this section, we test the generalization ability of the proposed HUFNO model for unseen hill shapes. The shape factors $a$ (cf. Eq. \ref{eq:hillshape}) in the training set are chosen at $a=1.0, 1.2, 1.4, 1.6, 1.8$, each of which contains 10 different initial conditions. Hence, the training set contains 50 simulation results in total. The other configurations remain the same as in the test for varying initial conditions. This test is conducted at $Re=700$ for demonstration. In the \emph{a posteriori} LES tests, we choose the unseen shape factors $a=0.9, 1.5$ and $1.9$, where $0.9$ represents a hill slope steeper than all those training ones, $1.5$ represents a interpolated shape, and $1.9$ represents a milder slope. The predicted turbulence statistics for different hill shapes are displayed in Fig. \ref{fig_compareles_velstat_varyshape}. The corresponding predictions by the traditional SMAG and WALE models are also shown in the figure. As can be seen, at all three new hill shapes, the HUFNO model performs better than the SMAG and WALE models, both of which show some discrepancies from the DNS benchmark, especially in the predicted Reynolds stresses.

\begin{figure}\centering
\includegraphics[height=.24\textwidth]{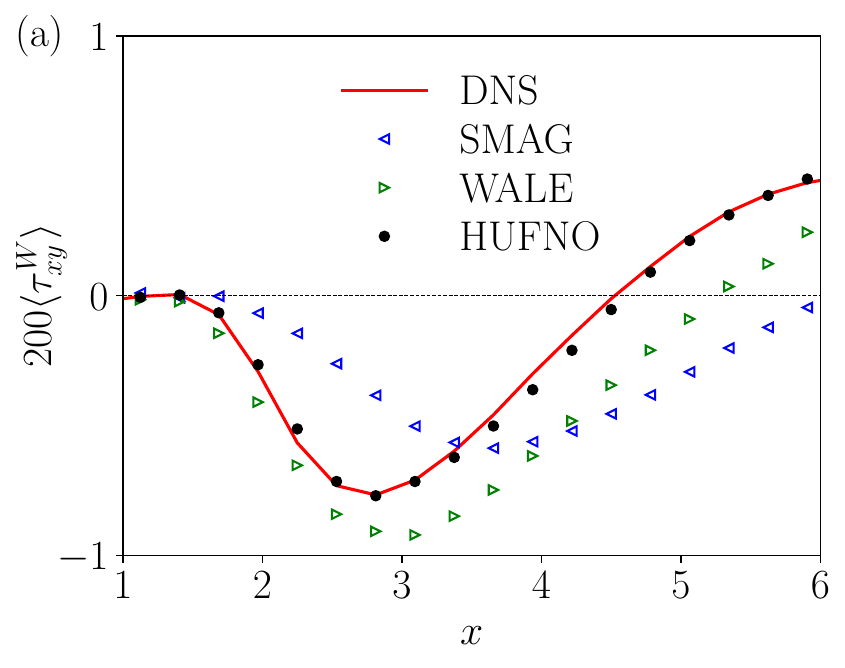}\hspace{-0.06in}
\includegraphics[height=.24\textwidth]{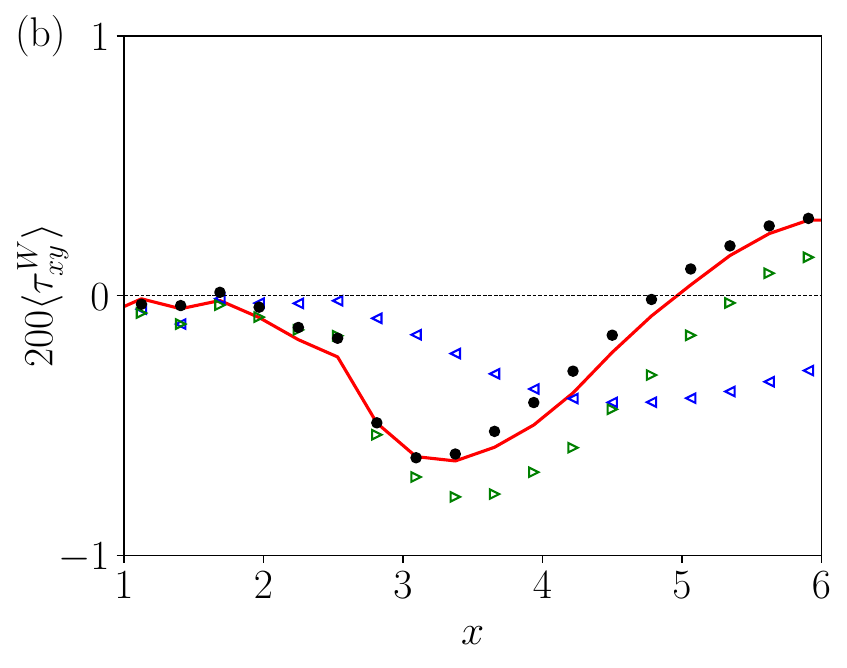}\hspace{-0.06in}
\includegraphics[height=.24\textwidth]{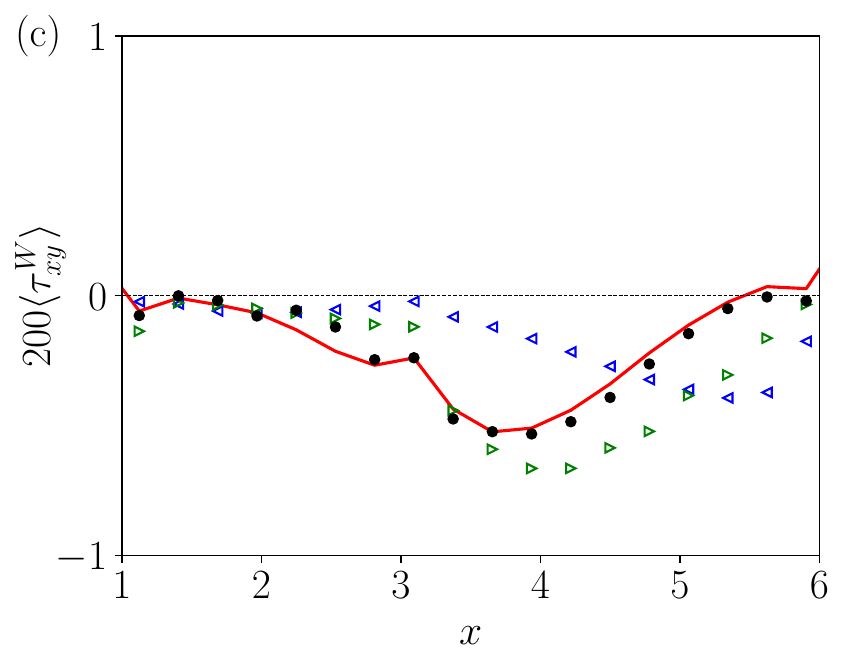}\vspace{-0.08in}
 \caption{The wall-shear stress in the flow separation region for varying hill shapes: (a) $a=0.9$; (b) $a=1.5$; (c) $a=1.9$.}\label{fig_compareles_tauw_shape}
\end{figure}

Undoubtedly, the change of the hill shape would affect the wall shear stress as displayed in Fig. \ref{fig_compareles_tauw_shape}. we observe that the separation zone extends farther downstream as the slopes become milder, indicating the strong influence of the hill shape on flow separation. Again, the WALE model tangibly outperforms the SMAG model due to its near-wall considerations in the formulation \cite{Nicoud1999}. Overall, the HUFNO model gives the best prediction of the wall shear stress at all three hill shapes, even though the prediction accuracy is slightly decreased compared to the scenario when only the initial condition is varied (cf. Fig. \ref{fig_compareles_tauw}). This is not surprising since generalizing to different boundary shapes is inherently more challenging than generalizing to different initial conditions. These results underscore the HUFNO's generalization ability in predicting turbulent flows over periodic hills with unseen shapes.

\subsubsection{\label{subsubsec:unseen_Re}Test at unseen Reynolds number}
%% Inline mathematics is tagged between $ symbols.

In this section, we study the performance of the HUFNO model at an unseen Reynolds number, which is important for fluid mechanics problem. In this test, the model is trained using data from $Re=700$ and $5600$, and ten groups of DNS datasets are generated using different random initial fluctuations at each Reynolds number for training. Then the \emph{a posteriori} test is performed at $Re=1400$. The grid resolution is invariably set at $48\times49\times16$ in the training and the \emph{a posteriori} test.

\begin{figure}\centering
\includegraphics[width=.3\textwidth]{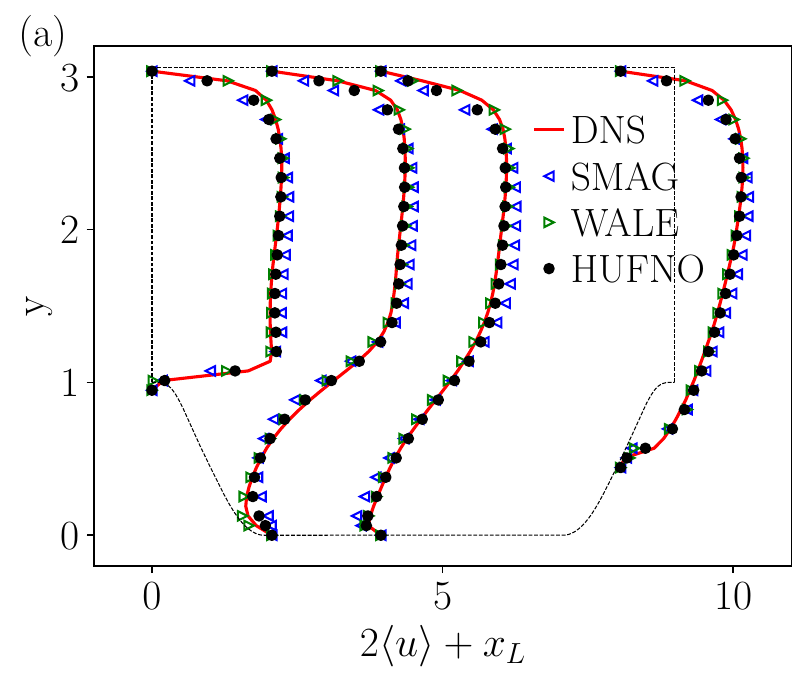}\hspace{-0.06in}
\includegraphics[width=.3\textwidth]{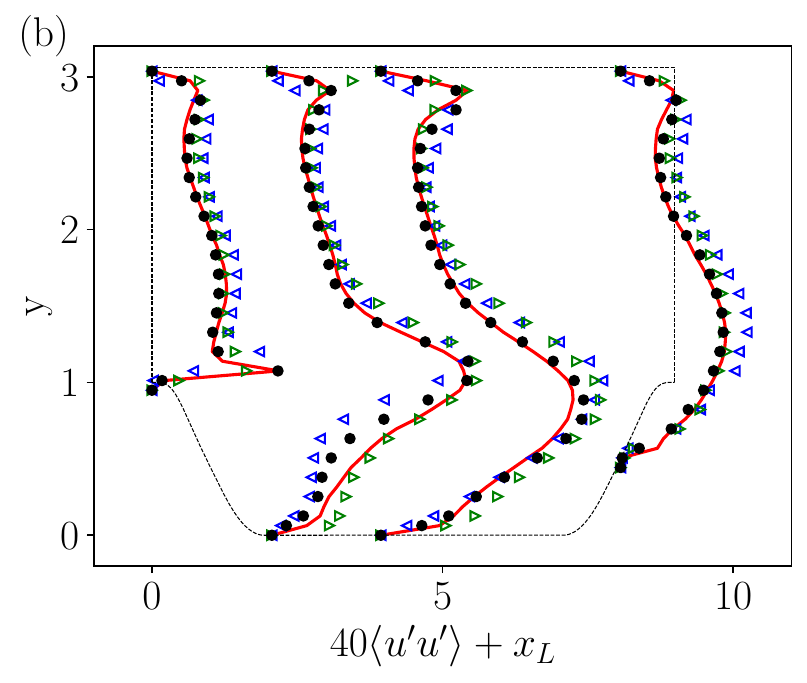}\hspace{-0.06in}
\includegraphics[width=.3\textwidth]{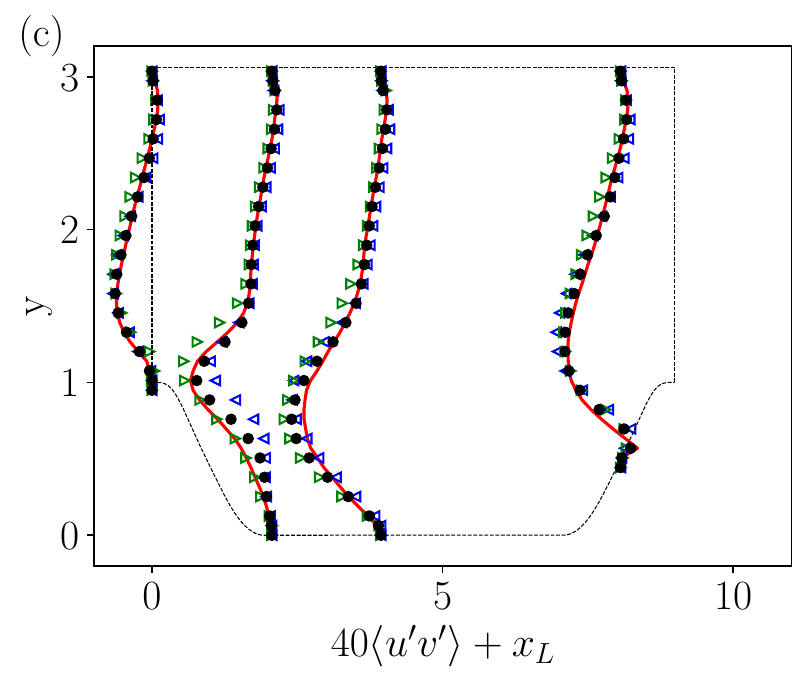}\vspace{-0.06in}

 \caption{The predicted turbulence statistics for turbulent flows over periodic hills at $Re=1400$, with the HUFNO model trained using the data at $Re=700$ and $5600$: (a) mean velocity $\langle u \rangle$; (b) normal Reynolds stress $\langle u'u' \rangle$; (c) shear Reynolds stress $\langle u'v' \rangle$.}\label{fig_compare_unseen_Re}
\end{figure}

The predicted turbulence statistics at $Re=1400$ is displayed in Fig. \ref{fig_compare_unseen_Re}, along with the predictions by traditional LES models. We observe that, without using the data from $Re=1400$ in the training process, reasonable agreement with the DNS results can still be achieved by the HUFNO model, underscoring its generalization ability at unseen Reynolds numbers. As the figure shows, even though the agreement with the DNS data is slightly worse than the case when data at $Re=1400$ is used (cf. Fig. \ref{fig_compareles_velstat}), the HUFNO model still outperforms the SM models while its accuracy is roughly similar to that of the WALE model. Nevertheless, we note that generalizing to unseen Reynolds numbers is inherently more difficult than generalizing to different initial conditions, and future research are still needed to further enhance the model's generalization abilities at more challenging Reynolds numbers.

\subsubsection{\label{subsubsec:3d_shape}Test in the case of three-dimensional hills}
%% Inline mathematics is tagged between $ symbols.

\begin{figure}\centering
\includegraphics[width=.5\textwidth]{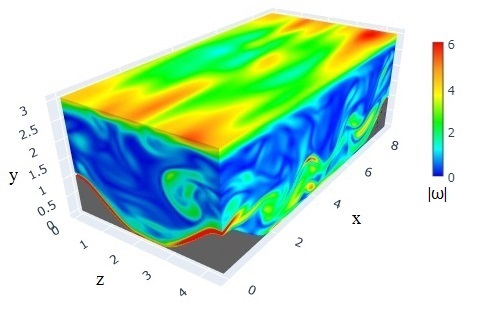}\hspace{-0.08in}
 \caption{The vorticity field of turbulent flows over three-dimensional hills.}\label{fig_3dhill}
\end{figure}

In this section, to further study the performance of the HUFNO model in a more complex geometry setting within the scope of separated turbulence, we test its performance on the turbulent flow over three-dimensional periodic hills (i.e. the shape of the hill varies in both the streamwise and spanwise directions like real hills). As illustrated in Fig. \ref{fig_3dhill}, the shape of the hill is like a circular dune with the shape factor $a=1.0$ while the mean flow is in the streamwise direction. The Reynolds number $Re=700$, and the grid resolution is $32\times33\times16$.

\begin{figure}\centering
\includegraphics[width=.3\textwidth]{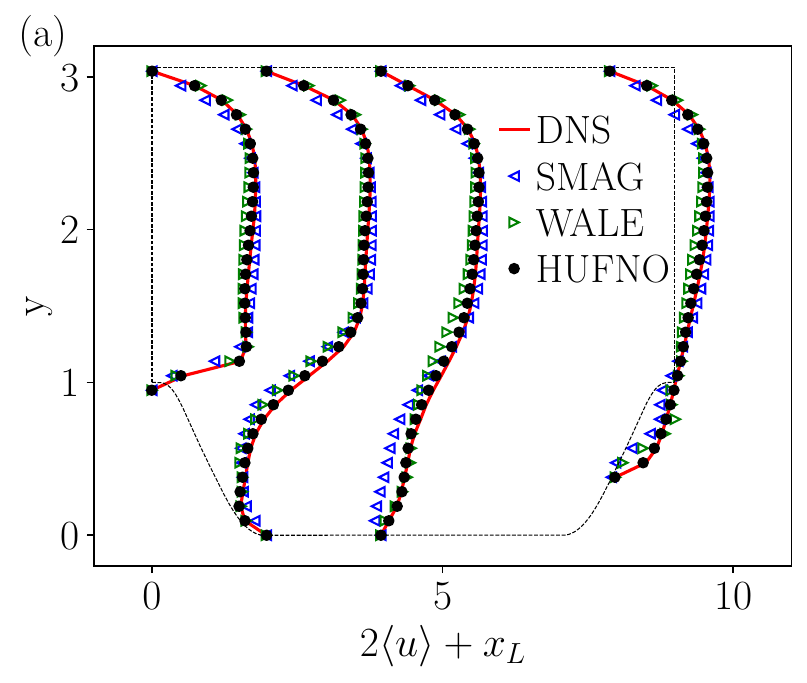}\hspace{-0.06in}
\includegraphics[width=.3\textwidth]{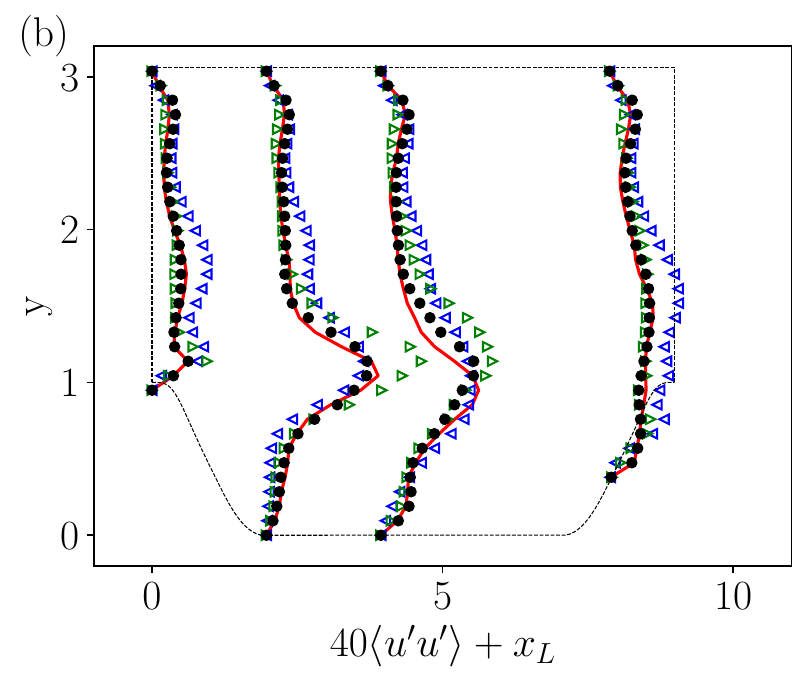}\hspace{-0.06in}
\includegraphics[width=.3\textwidth]{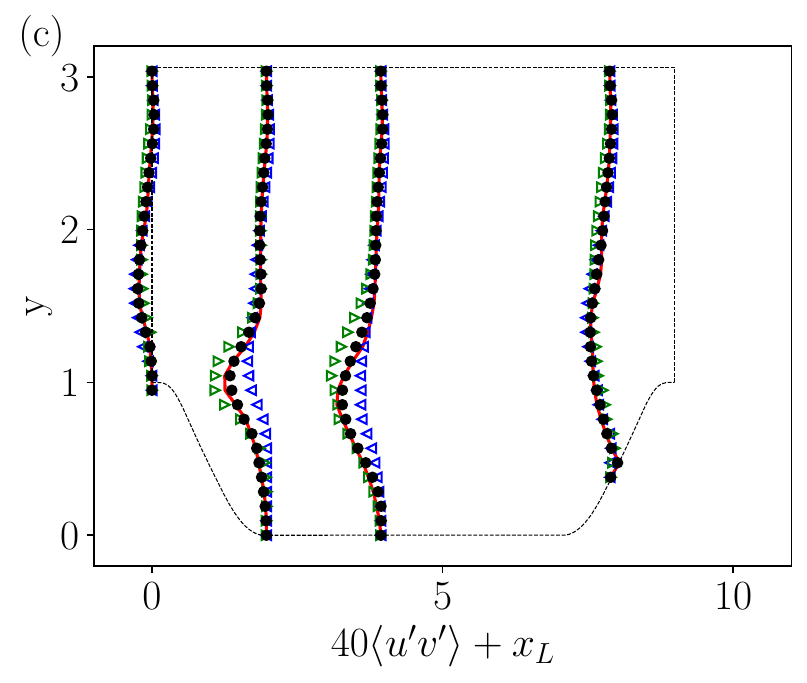}\vspace{-0.06in}

\includegraphics[width=.3\textwidth]{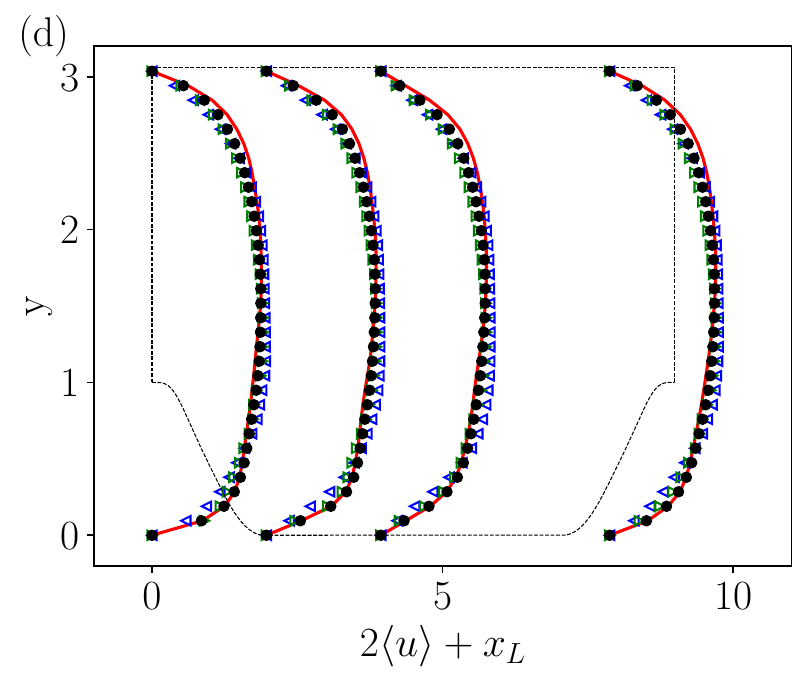}\hspace{-0.06in}
\includegraphics[width=.3\textwidth]{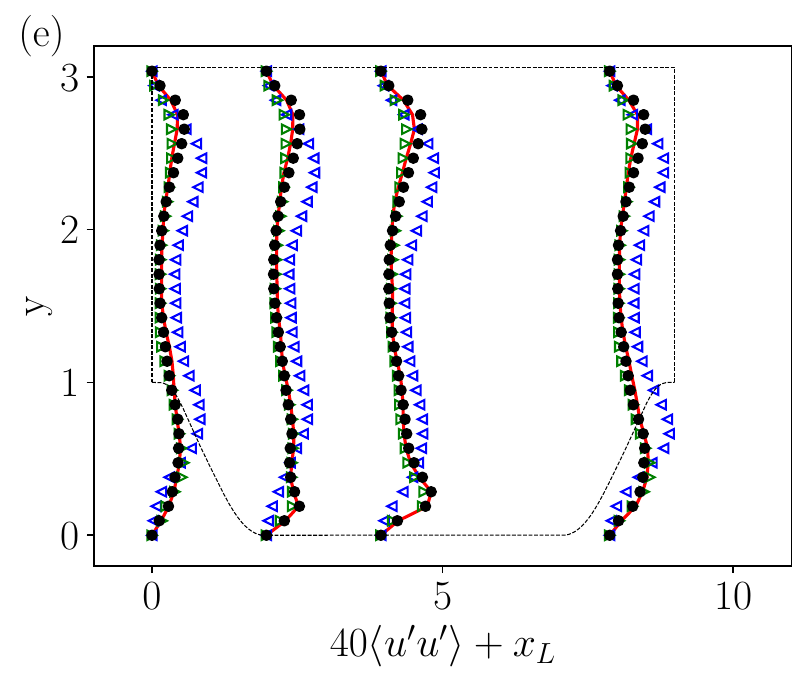}\hspace{-0.06in}
\includegraphics[width=.3\textwidth]{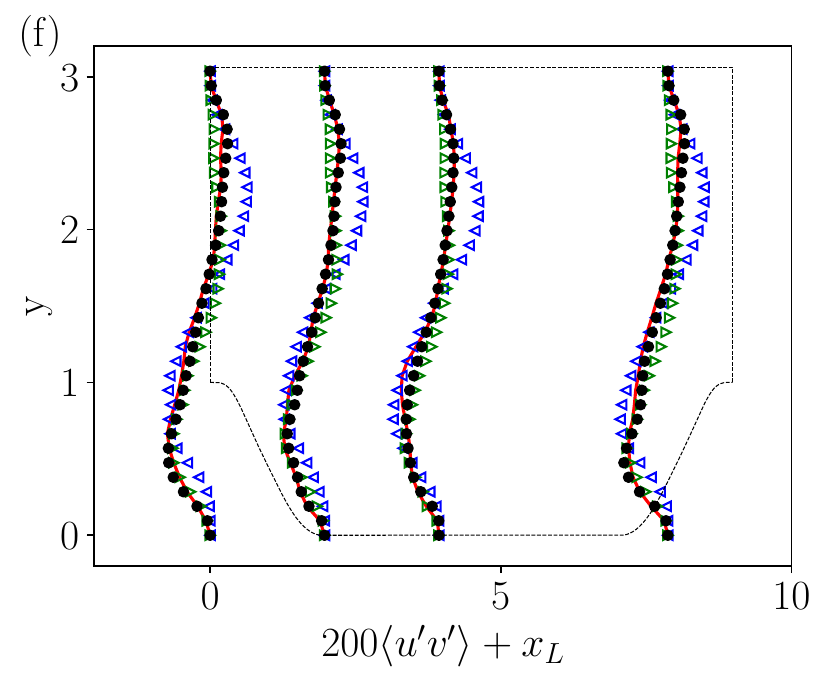}\vspace{-0.06in}

 \caption{The predicted turbulence statistics for turbulent flows over three-dimensional periodic hills: the first and second rows represent the results at the $x$-$y$ planes with $z=0$ and $z=L_z/2$, respectively; the first, second and third columns represent $\langle u \rangle$, $\langle u'u' \rangle$ and $\langle u'v' \rangle$, respectively.}\label{fig_compare_3d}
\end{figure}

The predicted turbulence statistics by HUFNO at different spanwise locations in the LES are shown in Fig. \ref{fig_compare_3d}, along with the predictions by traditional LES models. As can be seen, at the $x$-$y$ plane that cuts through the axis of the hills ($z=0$), the profiles of the velocity statistics strongly depend on the streamwise locations due to the presence of the hills. However, at the center $x$-$y$ plane between the two hills ($z=L_z/2$), the profiles of the velocity statistics are quite similar at different streamwise locations, even though they are not completely identical due to the influence of the hills. Overall, the HUFNO model performs better compared to the SMAG and WALE models, and matches reasonably well with the DNS results. This further demonstrate the capability of the HUFNO model.

%%%%%%%%%%%%%%%%%%%%%%%%%%%%%%%%%%%%%%%%%%%%%%%%%%%%%%%%%%%%%%%%%%%%%%%%%%%%%%%%%%%%%%%%%%%%%%%%%%%%%%%%%%%%%%%%%%%%%%%%%%%%%%%%%%%%%%%%%%%%%%%%%%%%%
\section{\label{sec:conclu}Conclusions}

In this study, a hybrid framework of U-Net and Fourier neural operator (HUFNO) is proposed and applied in the large-eddy simulations (LES) of turbulent flows over periodic hills. By integrating the advantages of the U-Net and Fourier neural operator (FNO), the proposed model is well suited for problems with mixed periodic and non-periodic boundary conditions. Our preliminary test shows that the HUFNO model outperforms the original FNO model and the U-Net model in the predictions of the turbulence statistics. In the \emph{a posteriori} tests of LES, the flow prediction capacity of the HUFNO framework is evaluated against the DNS benchmark and the traditional LES. 

The HUFNO framework is tested in four scenarios: at initial conditions different from the training set with fixed hill shape; at unseen hill shapes; at unseen Reynolds number; and in the case of three-dimensional hills. Compared to the traditional SMAG and WALE models-based LES, the newly proposed HUFNO model can predict better the mean velocity, the Reynolds stress, the energy spectrum, the invariant characteristics of velocity gradients and the wall-shear stresses. Meanwhile, the WALE model tangibly outperforms the SMAG model due to its inherent near-wall considerations in the formulation.

The streamlines and velocity contour of the flow field are examined to assess the ability of the HUFNO model in predicting the flow separation structures. Both the separation bubbles and re-circulation streamlines can be more accurately captured by the HUFNO model compared to the traditional LES models. Further, the computational costs of the HUFNO models are much lower than the traditional LES, demonstrating its great efficiency.

For potential applications, we note that the current HUFNO model is naturally suited for simulating 3D turbulent flows over bluffed bodies or arbitrary topographic features, such as flow around mountainous terrain \cite{Jing2020}, flow over dams \cite{Boulange2021}, flow over patterned geometries like urban microclimate-related flows \cite{Peng2024}. A key commonality of these scenarios is the frequent presence of non-periodic boundary conditions in the cross-stream directions, which the HUFNO framework is specifically designed to address.

Finally, we note that while the current work demonstrates the potential of HUFNO in the prediction of periodic hill turbulence, challenges still exist for machine learning-based flow solvers, including complex turbulent flows at high Reynolds numbers, irregular computational grids, and insufficient training data, etc. For example, insufficient training data may fail to cover diverse flow regimes (especially at high Reynolds numbers), which can lead to degraded model performance or long-term instabilities. Meanwhile, the performance of neural operator-based models such as HUFNO is also sensitive to grid resolution (similar to traditional numerical solvers) and hill geometry. That is, overly coarse grids may suppress small-scale flow details that are critical for capturing important physical processes such as energy dissipation, and extremely steep and irregular hill shapes may be outside the valid extrapolation range of the current model. Additionally, irregular and non-Cartesian grids are common in real engineering applications, and further research is still needed to extend the HUFNO to real-world geometries beyond periodic hills.

To address these difficulties, many aspects can be considered in future work. For example, incorporating physical constraints can potentially lower data requirements \cite{Raissi2019,Wang2023,Li2023d,Zhao2025}, as embedded physical knowledge may compensate for limited training samples. Incorporating geometry-informed neural operators (GINO) can help handle complex geometries while adapting to irregular grids and extreme shapes \cite{Li2023c}, thereby overcoming the limitations of the current HUFNO framework. Integrating data assimilation techniques to fuse sparse real-world observational data with model predictions also has the potential to further correct model errors resulting from the aforementioned limitations \cite{He2020,Mons2021}.

\begin{acknowledgments}
This work was supported by the National Natural Science Foundation of China (NSFC Grants No. 12302283, No. 12588301, and No. 12172161), by NSFC Excellence Research Group Program for `Multiscale Problems in Nonlinear Mechanics' (No. 12588201), by the Shenzhen Science and Technology Program (Grant Nos. SYSPG20241211173725008, and KQTD20180411143441009), by Department of Science and Technology of Guangdong Province (Grants No. 2019B21203001, No. 2020B1212030001, and No.2023B1212060001), by the Shandong Province Youth Fund Project (Grant No. ZR20240A050), and by the Natural Science Foundation of Hainan Province (Grant No. 425QN376). Additional support was provided by the Innovation Capability Support Program of Shaanxi (Program No. 2023-CX-TD-30). This work was also supported by Center for Computational Science and Engineering of Southern University of Science and Technology.
\end{acknowledgments}
%\nocite{*}

\end{document}